\documentclass[manuscript]{revtex4}
\usepackage{cancel}
\usepackage{shuffle}
\usepackage{draft}
\usepackage{amsfonts}
\usepackage{mathrsfs}
\usepackage{amsmath}
\usepackage{amssymb}
\usepackage{bm}
\usepackage{feynmp-auto}
\usepackage[normalem]{ulem}
\usepackage{extarrows}
\usepackage{slashed}
\usepackage{float}
\usepackage{isodateo}
\usepackage{graphicx}
\usepackage{xcolor}
\usepackage[CJKbookmarks=true, bookmarksnumbered=true,bookmarksopen=true]{hyperref}
\hypersetup{colorlinks,%
	linkcolor=[rgb]{0,0.2,0.6}, %
	urlcolor=[rgb]{0,0.2,0.6}, %
	citecolor=[rgb]{0,0.2,0.6}}

\usepackage{indentfirst}

\graphicspath{{fig/}}
\usepackage{bm}
\usepackage{indentfirst}
\usepackage{tikz-cd}
\usetikzlibrary{patterns}
\usepackage{tikz}
\usetikzlibrary{arrows,shapes}
\usetikzlibrary{trees}
\usetikzlibrary{matrix,arrows} 				
\usetikzlibrary{positioning}				
\usetikzlibrary{calc,through}				
\usetikzlibrary{decorations.pathreplacing}  
\usepackage{pgffor}							

\usetikzlibrary{decorations.pathmorphing}	
\usetikzlibrary{decorations.markings}
\tikzset{
	photon/.style={decorate, decoration={snake}, draw=red},
	electron/.style={draw=blue, postaction={decorate},
		decoration={markings,mark=at position .55 with {\arrow[draw=blue]{>}}}},
	gluon/.style={decorate, draw=blue,
		decoration={coil,amplitude=4pt, segment length=4pt}} ,
	vector/.style={decorate, decoration={snake}, draw},
	provector/.style={decorate, decoration={snake,amplitude=2.5pt}, draw},
	antivector/.style={decorate, decoration={snake,amplitude=-2.5pt}, draw},
	fermion/.style={draw=black, postaction={decorate},
		decoration={markings,mark=at position .55 with {\arrow[draw=black]{>}}}},
	fermionbar/.style={draw=black, postaction={decorate},
		decoration={markings,mark=at position .55 with {\arrow[draw=black]{<}}}},
	fermionnoarrow/.style={draw=black},
	scalar/.style={dashed,draw=black, postaction={decorate},
		decoration={markings,mark=at position .55 with {\arrow[draw=black]{>}}}},
	scalarbar/.style={dashed,draw=black, postaction={decorate},
		decoration={markings,mark=at position .55 with {\arrow[draw=black]{<}}}},
	scalarnoarrow/.style={dashed,draw=black},
	electron/.style={draw=black, postaction={decorate},
		decoration={markings,mark=at position .55 with {\arrow[draw=black]{>}}}},
	bigvector/.style={decorate, decoration={snake,amplitude=4pt}, draw},
	background/.style={dashed,draw=black, postaction={decorate},
		decoration={markings,mark=at position 1 with {\arrow[draw=black]{<>}}}},
}

\tikzstyle{block} = [draw, rectangle, 
minimum height=3em, minimum width=6em]

\def\mb{\mathbf}
\def\ma{\mathcal}

\usepackage{mdframed}

\newmdenv[skipabove=0pt,%
skipbelow=5pt,%
leftmargin=0pt,%
rightmargin=0pt,%
innertopmargin=-5pt,%
innerbottommargin=7pt,%
innerleftmargin=2pt,%
innerrightmargin=2pt,%
splittopskip=0pt,%
splitbottomskip=0pt,%
linewidth=0pt,%
nobreak=true]%
{keyeqn2} 
\newmdenv[backgroundcolor=gray!15,%
skipabove=0pt,%
skipbelow=5pt,%
leftmargin=0pt,%
rightmargin=0pt,%
innertopmargin=-5pt,%
innerbottommargin=7pt,%
innerleftmargin=2pt,%
innerrightmargin=2pt,%
splittopskip=0pt,%
splitbottomskip=0pt,%
linewidth=0pt,%
nobreak=true]%
{keyeqn}

\begin{document}
\title{Differential Operators and Unifying Relations for 1-loop Feynman Integrands from Berends-Giele Currents}
	
\author{Qi Chen}
\email{chenq20@mails.tsinghua.edu.cn}
\affiliation{Department of Physics, Tsinghua University, Beijing 100084, China}
\author{Yi-Xiao Tao}
\email{taoyx21@mails.tsinghua.edu.cn}
\affiliation{Department of Mathematical Sciences, Tsinghua University, Beijing 100084, China}

\begin{abstract}
Our work focuses on utilizing the Berends-Giele currents to construct differential operators and unifying relations for 1-loop Feynman integrands. We successfully reproduce the known results for the unifying relations between Yang-Mills theory and Yang-Mills scalar theory, and extend the discussion to the (A)dS case for the scalar theory with minimal coupling to gluons.
\end{abstract}

\maketitle

\tableofcontents	
\newpage	
\section{Introduction}

Unifying relations, as described in \cite{Cheung:2017ems}, reveal the implicit connections between amplitudes in various theories. These relations can be proven through on-shell factorization or by utilizing the Cachazo-He-Yuan (CHY) formalism \cite{Cachazo:2013hca,Cachazo:2013iea,Cachazo:2014xea}, which involves applying operators to CHY integrands \cite{Zhou:2018wvn}. By establishing a unifying web that interconnects many theories, these relations also aid in comprehending the universal expansion of tree-level amplitudes \cite{Zhou:2019mbe,Dong:2021qai}. A recent work \cite{Tao:2022nqc} has discovered another proof of the unifying relations. In this work, a corollary of the original relations is demonstrated using Berends-Giele (BG) recursions \cite{Berends:1987me} in (A)dS spacetime, which can also be applied to the flat spacetime. BG recursion has its advantages and disadvantages when compared to on-shell factorization. It is a semi-on-shell method with its origins in Feynman rules, making it a straightforward and intuitive approach to generalizing on-shell concepts to off-shell situations. However, BG currents are gauge-dependent, and the chosen gauge must be kept in mind during the calculation of BG currents. It should be noted that several other recursion methods exist for AdS spacetime \cite{Raju:2010by,Raju:2011mp,Raju:2012zr,Albayrak:2019asr} as well as dS spacetime \cite{Baumann:2019oyu,Baumann:2020dch,Baumann:2020ksv,Baumann:2022jpr,Arkani-Hamed:2015bza,Arkani-Hamed:2018kmz}.


A recent study by \cite{Gomez:2022dzk} showed that BG currents can also be utilized to construct 1-loop integrands, which differs from the traditional on-shell methods\cite{Arkani-Hamed:2010zjl,Boels:2010nw,Farrow:2020voh,Cachazo:2015aol}. To be precise, we can first take all the legs of the BG currents off-shell and then ``sew" two off-shell legs together to obtain a loop integrand. This approach enables us to generalize the proof of the unifying relations from the tree level to the 1-loop level. Previously, the proof at the 1-loop level relied on the so-called ``forward limit" \cite{He:2015yua}, which can join two on-shell legs to form a loop in the CHY formalism and obtain the corresponding differential operators for the 1-loop integrand \cite{Zhou:2021kzv}. Another method to the off-shell recursion relation at the loop level has been presented in \cite{Lee:2022aiu}, based on the Dyson–Schwinger equation. This paper mainly focuses on the unifying relations between the Yang-Mills (YM) theory and the Yang-Mills scalar (YMS) theory (we will give more details of the YMS theory in Section \ref{sec:flatproof}) at the 1-loop level and reproduces the differential operators for 1-loop integrands using BG currents. After discussing the flat case, we generalize the sewing procedure to the (A)dS case and explore the relationship between the YM theory and the conformally coupling scalar theory with gluon minimal coupling, which is of greater significance in cosmology. An alternate approach to loop amplitudes in AdS can be found in \cite{Albayrak:2020bso}.


The structure of this paper is as follows. In Section \ref{sec:flatprooftree}, we provide a brief overview of the proof of the unifying relations at the tree level using BG currents. In Section \ref{sec:flat}, we discuss the procedure of sewing two off-shell legs of the off-shell BG currents. Section \ref{sec:flatproof} is devoted to the proof of the unifying relations at the 1-loop level. We begin by demonstrating how to obtain differential operators from the sewing procedure through a simple example, i.e., the scalar theory with a gluon minimal coupling. Next, we present a brief proof of the unifying relations between the YM theory and the YMS theory at the 1-loop level, based on BG recursions and the construction of the multi-trace YMS currents. In Section \ref{sec:ads}, we extend the sewing procedure to (A)dS and investigate the unifying relations in this context.

\section{Differential Operators and Unifying relations at the tree Level in Flat Spacetime}\label{sec:flatprooftree}

In this section, we will revisit some elegant unifying relations that unveil intrinsic connections between seemingly different theories. Specifically, we will concentrate on a particular unifying relation, which can be seen as a simplified consequence of the unifying relation between YM and YMS theories, where the scalar is minimally coupled to gluons. Additionally, we will present an off-shell approach to establish the unifying relation between pure YM and scalars minimally coupled to gluons. It is important to note that all discussions in this section will take place in Minkowski spacetime.

\subsection{Review on Unifying Relations}
Although the Lagrangian formulation suggests that the YMS theory differs from pure YM theory, they are intrinsically connected at the tree level. Specifically, differential operators can transform the amplitudes of pure YM theory into those of YMS theory, which are determined by on-shell kinematics and gauge invariance \cite{Cheung:2017ems,Zhou:2018wvn,Bollmann:2018edb}. Recent studies have also shown that these differential operators can be generalized to 1-loop amplitudes \cite{Zhou:2021kzv,Zhou:2022djx}. It should be noted that the YM and YMS theory, as well as the bi-adjoint scalar (BS) theory and even gravity theory, can exhibit connections in terms of amplitudes. However, this paper focuses mainly on the unifying relation between YM and YMS.  

Let's express the unifying relation between YM and YMS in a clearer way. Color-ordered amplitudes for YM and YMS are related by the following equation:
\begin{equation}\label{eq:unifyingYMSandYM}
	A_{\rm YMS\,}=\mathcal{T}[i_{1}j_{1}]\mathcal{T}[i_{2}j_{2}]\cdots\mathcal{T}[i_{n}j_{n}]\cdot A_{\rm YM\,}
\end{equation}
where $\ma{T}[ij]$ is the differential operator transferring the color-ordered amplitude from one theory to another theory. More precisely, it is defined as:
\begin{equation}
	\mathcal{T}[{ij}]=\partial_{\epsilon_{i}\epsilon_{j}},
\end{equation}
where $\epsilon_i$ is the polarization vector of the $i$-th particle. These differential operators are also referred to as ``trace operators". We should note that the particles denoted by indices in each differential operator $\mathcal{T}[\alpha]$ are in the same trace after $\mathcal{T}[\alpha]$ acts on a certain amplitude.  For $|\alpha|\geq2$, the operator $\ma{T}[\alpha]$ can be expressed as:
\ie
\mathcal{T}[\alpha]=\mathcal{T}_{\alpha_{1}\alpha_{n}}\cdot\prod_{i=2}^{n-1}\mathcal{T}_{\alpha_{i-1}\alpha_{i}\alpha_{n}},
\fe
where $\ma{T}_{ij}=\ma{T}[ij]$ and $\mathcal{T}_{ijk}=\partial_{k_{i}\epsilon_{j}}-\partial_{k_{k}\epsilon_{j}}$ with $k_{i}$ being the momentum of the $i$-th particle. It is worth noting that there can be connections between amplitudes in other theories as well, such as the bi-adjoint scalar (BS) theory and gravity theory, but this paper primarily focuses on the unifying relation between YM and YMS.


It is noteworthy that the unifying relation \eqref{eq:unifyingYMSandYM} between YM and YMS theory can be extended to encompass partial trace operators. By partial trace operators, we mean that it is possible to construct differential operators for a subset of the external legs rather than for all the legs. Thus, the partial trace operators can transfer only some of the gluons in pure YM amplitudes to scalars and yield a mixed amplitude with both gluons and scalars. For instance, let us consider a 4-point gluon amplitude $A_{\rm YM}(g_{1},g_{2},g_{3},g_{4})$. Then, we can apply the partial trace operator $\ma{T}[123]$ to obtain a mixed amplitude $A_{\rm YMS}(\phi_{1}\phi_{2}\phi_{3},g_{4})$:
\begin{equation}\label{eq:unifyingYMSandYMmix}
	A_{\rm YMS}(\phi_{1}\phi_{2}\phi_{3},g_{4})=\mathcal{T}[123]\cdot A_{\rm YM}(g_{1},g_{2},g_{3},g_{4}).
\end{equation}

The unifying relations \eqref{eq:unifyingYMSandYM} establish a connection between the YM theory and the YMS theory. This connection can be extended to the scalar theory with minimal coupling to gluons, where $\phi^{3}$ and $\phi^{4}$ vertices do not exist, and scalars have only one color index. In this case, the second color group of the YM scalar is chosen to be the $U(1)$ group, and the scalars separated from other scalars by gluon propagators are in the same trace. Obviously, each trace can only have two scalars. Therefore, all possible trace structures must be summed over to obtain the color-ordered amplitudes. To establish the connection between the minimally coupled scalar and YM theory at amplitudes, the unifying relations \eqref{eq:unifyingYMSandYM} can be modified as follows:
\ie\label{eq:unifyingminicouplescalarandYM}
\ma{T}^{X}A_{\rm YM}(g_{1},g_{2}\cdots g_{n})=A_{\rm S}(\phi_{X},g_{{1,2,\cdots,n}\backslash X})
\fe
Here, $A_{\rm S}$ is the amplitude of the scalar theory with minimal coupling to gluons, and $\phi_{X}$ denotes the letters in $X\subseteq \{1,2,3,\cdots,n\}$ that correspond to scalar legs in $A_{\rm S}$, while $g_{{1,2,\cdots,n}\backslash X}$ correspond to gluon legs (here the word $X$ is only used to label which particles are scalars, and the order of the particles in $A_{\rm S}$ is the same as in $A_{\rm YM}$). The operator $\ma{T}^{X}$ pairs the letters in the word $X$ to obtain a product of $\ma{T}[ij]$ based on this pairing method and then sums over all pairing methods. For example, $\ma{T}^{1234}=\ma{T}[12]\ma{T}[34]+\ma{T}[13]\ma{T}[24]+\ma{T}[14]\ma{T}[23]$. The length of the word $X$ must be even since the theory cannot have an amplitude with an odd number of scalar legs. The connection between the minimally coupled scalar and YM theory at amplitudes is also referred to as unifying relations.
 
\subsection{Semi-on-shell Methods Approach to Unifying Relations}

In prior work, the unifying relations between YM and YMS \eqref{eq:unifyingYMSandYM} were established via on-shell recursive methods \cite{Cheung:2017ems} and were shown to be consistent with CHY formulae \cite{Zhou:2018wvn,Bollmann:2018edb}. While the factorization and on-shell methods have been well-developed in flat spacetime, the study of such factorization methods in curved spacetime, especially in dS spacetime, remains incomplete. Therefore, in order to extend the unifying relation \eqref{eq:unifyingYMSandYM} to curved spacetime, we introduce a method that can avoid the intricacies of factorization in curved spacetime. The BG currents recursion method is notable for its semi-on-shell (semi-on-boundary in (A)dS) property \cite{Berends:1987me,Armstrong:2022jsa,Tao:2022nqc}. In this subsection, we examine the proof of \eqref{eq:unifyingminicouplescalarandYM} using the BG currents, taking the unifying relations between minimally coupled scalar and YM theory as an example.

The currents in the $n$-pt BG can be established by utilizing the $n$-pt tree level correlation functions, with one external leg being off-shell. Additionally, these BG currents satisfy the non-linear equation of motion, as stated in references \cite{Lee:2015upy} and \cite{Mafra:2016ltu}. The perturbiner formalism can be employed systematically to approach these currents \cite{Selivanov:1999as}. For a more detailed demonstration of this procedure, we can consider the YM theory. The equation of motion for YM theory takes the form:
\begin{equation}\label{eq:gluoneomflatnoghost}
	\square \mb{A}_{\mu}=-i[\mb{A}^{\nu},\mb{F}_{\mu\nu}]+i[\mb{A}^{\nu},\partial_{\nu}\mb{A}_{\mu}].
\end{equation}
Here, $\mb{A}_{\mu}=A_{\mu}^aT^a$ denotes the Lie algebra-valued gluon field, with $T^a$ being the Lie group generators. The corresponding Lie algebra-valued field strength tensor is given by $\mb{F}_{\mu\nu}=\partial_{\mu}\mb{A}_{\nu}-\partial_{\nu}\mb{A}_{\mu}-i[\mb{A}_{\mu},\mb{A}_{\nu}]$. Note that the Lorenz gauge is assumed in this case. To obtain the multi-particle solution of \eqref{eq:gluoneomflatnoghost}, we can apply the perturbiner expansion ansatz, which proceeds as follows:
\begin{eqnarray}\label{eq:gluonansatznoghost}
    \mb{A}_{\mu}&=&\sum_P \ma{A}_{P\mu}e^{ik_P\cdot x}T^{P}\\
    \mb{F}^{\mu\nu}&=&i\sum_{P}\mathcal{F}^{\mu\nu}_{P}T^{P}e^{ik_{P}x}
\end{eqnarray}
where $P$ is a bunch of letters $I=i_1\cdots i_s$ and each $i_{s}$ represents for a single particle state. Here $k_{P\mu}=k_{1\mu}+\cdots+k_{s\mu}$ is the total momentum for the multi-particle state and $T^{a_P}=T^{a_{i_1}}\cdots T^{a_{i_s}}$. 
The Lie algebra valued $\mb{A}_{\mu}$, which is in the adjoint representation of some gauge group, must satisfy the shuffle identity\cite{Kleiss:1988ne,Lee:2015upy,Mizera:2018jbh}:
\begin{eqnarray}\label{eq:shuffleidentity}
	\ma{A}_{P\shuffle Q}^{\mu}=\ma{F}_{P\shuffle Q}^{\mu\nu}=0,\ \ P,Q\neq\varnothing
\end{eqnarray}
Here, $\shuffle$ indicates that we sum over all permutations of the labels in $P\cup Q$ while preserving the ordering of the labels in $P$ and $Q$. From the shuffle identity \eqref{eq:shuffleidentity}, the perturbiner expansion ansatz \eqref{eq:gluonansatznoghost} can be written as a sum of Lie brackets of all orders (i.e. a sum of $[T^{a_{1}}, T^{a_{2}}]$, $[T^{a_{1}},[T^{a_{2}}, T^{a_{3}}]]$ and so on), which means we can regard this ansatz as a ``power series" of the structure constant $f^{abc}$ for some gauge group. The expansion coefficients $\ma{A}_{P\mu}$ and $\ma{F}_P^{\mu\nu}$ do not contain any color degrees of freedom. These color-stripped coefficients are the BG currents\cite{Berends:1987me,Lee:2015upy,Mizera:2018jbh}. By substituting the perturbiner expansion ansatz \eqref{eq:gluonansatznoghost} and shuffle identity \eqref{eq:shuffleidentity} into the equation of motion \eqref{eq:gluoneomflatnoghost}, we can obtain the BG currents for gluons:
\begin{eqnarray}\label{eq:gluoncurrentnoghostflat}
    -s_{P}\mathcal{A}_{P}^{\mu}&=&\sum_{P=XY}[\mathcal{A}_{X}^{\mu}(k_{X}\cdot \mathcal{A}_{Y})+\mathcal{A}_{X\nu}\mathcal{F}_{Y}^{\mu\nu}-(X\leftrightarrow Y)]\\
\mathcal{F}_{Y}^{\mu\nu}&=&k_{Y}^{\mu}\mathcal{A}_{Y}^{\nu}-k_{Y}^{\nu}\mathcal{A}_{Y}^{\mu}-\sum_{Y=RS}(\mathcal{A}_{R}^{\mu}\mathcal{A}_{S}^{\nu}-\mathcal{A}_{S}^{\mu}\mathcal{A}_{R}^{\nu}).
\end{eqnarray}
In the equation for the BG currents, $1/s_P$ denotes the propagator for the off-shell leg, and the deconcatenation sum $P=XY$ represents the sum over all possible order-preserving ways of splitting $P$ into $X$ and $Y$. For example, if $P=12345$, then $X=123$ and $Y=45$. It's worth noting that each deconcatenation can be represented by a binary tree diagram \cite{Frost:2020eoa}.

\begin{figure}
\centering
\begin{tikzpicture}[line width=1pt,scale=1.5]
\draw[fermionnoarrow](2,-1)--(3,0);
\draw[fermionnoarrow](1.5,-0.5)--(2,0);
\draw[fermionnoarrow](1.5,-0.5)--(1,0);
\draw[fermionnoarrow](2,-1)--(1.5,-0.5);
\draw[fermionnoarrow](2,-1)--(2,-1.5);
\node at (1,0.2) {1};
\node at (2,0.2) {2};
\node at (3,0.2) {3};
\node at (2,-2) {$(a)$};
\begin{scope}[shift={(4,0)}]
\draw[fermionnoarrow](2,-1)--(3,0);
\draw[fermionnoarrow](2.5,-0.5)--(2,0);
\draw[fermionnoarrow](1.5,-0.5)--(1,0);
\draw[fermionnoarrow](2,-1)--(1.5,-0.5);
\draw[fermionnoarrow](2,-1)--(2,-1.5);
\node at (1,0.2) {1};
\node at (2,0.2) {2};
\node at (3,0.2) {3};
\node at (2,-2) {$(b)$};
\end{scope}
\end{tikzpicture}
\caption{Binary representations for Lie monomials. $(a)$ is the diagrammatic representation for [[1,2],3] and $(b)$ represents the Lie algebra structure for [1,[2,3]].}
\label{fig:binarytree}
\end{figure}
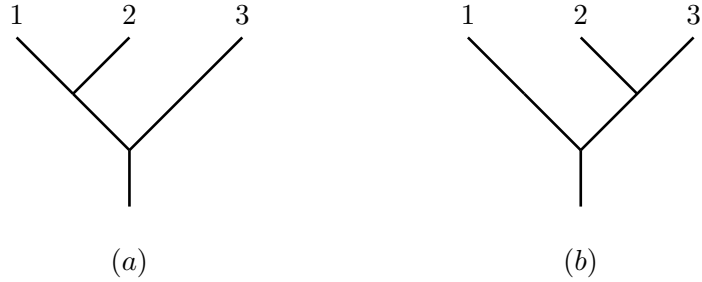

By substituting the perturbiner expansion ansatz \eqref{eq:gluonansatznoghost} and the shuffle identity \eqref{eq:shuffleidentity} into the equation of gluon motion \eqref{eq:gluoneomflatnoghost}, we can obtain Lie monomials such as $[[T^1,T^2],T^3]$ (for simplicity, we use $T^i$ as $i$, and denote the Lie monomial $[[T^1,T^2],T^3]$ as $[[1,2],3]$). Each Lie monomial corresponds to a binary tree, where $1$ and $2$ are two different branches that originate from the vertex, and $1$ and $2$ together with $3$ are two different branches from a new vertex. This binary tree can be diagrammatically represented as shown in Figure \ref{fig:binarytree} (a), and the deconcatenation $P=XY$ for this binary tree is $P=123$, $X=12$, $Y=3$, $R=1$, and $S=2$.

For the 3-point BG currents, another deconcatenation is possible: $P=123$, $X=1$, $Y=23$, $R=2$, and $S=3$. This deconcatenation can be diagrammatically represented by the binary tree shown in Figure \ref{fig:binarytree} (b). The full 3-point gluon BG current can be obtained from the sum of the two binary trees. More generally, for $n$-point BG currents, we can obtain the full currents by summing over all possible deconcatenations, and each deconcatenation corresponds to a unique binary tree, which can be thought of as a Lie monomial version of Feynman diagrams.

The recursive BG currents for minimally coupled scalars can be derived using the same approach as for the gluon case. To begin, we write down the equation of motion for a scalar field that is minimally coupled to gluons. When we turn on the interaction between gluons and scalars, the equation of motion for the gluons must also be modified:
%
\begin{eqnarray}
    \square \mb{A}_{\mu}&=&-i[\mb{A}^{\nu},\mb{F}_{\mu\nu}]+i[\mb{A}^{\nu},\partial_{\nu}\mb{A}_{\mu}]-\mb{J}_{\mu}\\
    \square\phi&=&i[\partial_{\mu}\phi,\mb{A}^{\mu}]-i[\mb{A}^{\mu},\partial_{\mu}\phi]+[[\mb{A}^{\mu},\phi],\mb{A}_{\mu}],\label{eq:scalareomminkowski}
\end{eqnarray}
The interaction current between the minimally coupled scalar and gluons is denoted by $\mb{J}_{\mu}=[(i\partial_{\mu}\phi+[\mb{A}_{\mu},\phi]),\phi]$, where we have chosen the Lorenz gauge. In the case of a scalar in the adjoint representation of the gauge group, the perturbiner expansion ansatz for multi-particle states can be arranged as follows:
%
\begin{equation}\label{eq:scalaransatzflat}
    \phi=\sum_I \phi_I(z)T^{a_I}e^{ik_I\cdot x}.
\end{equation}
By substituting the scalar perturbiner expansion ansatz \eqref{eq:scalaransatzflat} into the equation of motion \eqref{eq:scalareomminkowski}, we can derive the BG currents for the scalars that are minimally coupled to gluons:
\begin{equation}\label{eq:scalarrecurflat}
    \begin{split}
        -s_P\phi_I&=\sum_{I=JK}[2\phi_J(k_J\cdot\tilde{\ma{A}}_K)-(J\leftrightarrow K)]\\
        &+\sum_{I=JKL}[(\tilde{\ma{A}}_J\cdot\tilde{\ma{A}}_K)\phi_L-(\tilde{\ma{A}}_J\cdot\tilde{\ma{A}}_L)\phi_K+(J\leftrightarrow L)].
    \end{split}
\end{equation}
To distinguish the BG currents for pure YM theory and the gluons minimally coupled to scalars, we add a tilde to the latter. The propagator for scalars is the same as that for gluons as their equation of motion coincides at linear order. To obtain the gluon BG currents with the interaction, we add the interaction current to the BG currents we derived in the pure gauge case \eqref{eq:gluoncurrentnoghostflat}:
%
\begin{equation}
    \mathcal{J}_{I\mu}=\sum_{I=JK}(-k_{J\mu}\phi_{J}\phi_{K}+k_{K\mu}\phi_{K}\phi_{J})
+\sum_{I=JKL}[\tilde{\mathcal{A}}_{J\mu}\phi_{K}\phi_{L}-2\tilde{\mathcal{A}}_{K\mu}\phi_{J}\phi_{L}+\phi_{J}\tilde{\mathcal{A}}_{L\mu}\phi_{K}].
\end{equation}

The starting point for constructing the BG currents is the one-point current or the single particle state. For gluons, the initial condition is simply the single particle polarization function $\epsilon_i$. For scalars, the initial condition is the wave function $\phi_i$, which satisfies the Klein-Gordon equation. It is important to note that the $n$-point correlation function can be constructed from the $(n-1)$-point BG currents:
\begin{eqnarray}\label{eq:correlationfunctiondef}
    A_{\rm YM\,}(1,2,\cdots,N)&=&s_P\mathcal{A}_{N}\cdot\ma{A}_{1\cdots N-1},\\
    A_{\rm S\,}(1,2,\cdots,N)&=&s_P \phi_{N}\phi_{1\cdots N-1}.
\end{eqnarray}

In the following, we will outline the proof of the relation \eqref{eq:unifyingminicouplescalarandYM}. The detailed proof can be found in our previous work, where we have presented the complete derivation of the unifying relations \eqref{eq:unifyingminicouplescalarandYM} in (A)dS \cite{Tao:2022nqc}.

To begin with, let us reiterate that the deconcatenation sum in the BG currents can be represented as a unique binary tree. Thus, our proof will focus on an arbitrary given binary tree. We will demonstrate that even for an arbitrary binary tree, the BG current for gluons and minimally coupled scalars can be related by differential operators. These differential operators can also connect the total BG currents for these two theories. Finally, by invoking the definition of correlation function \eqref{eq:correlationfunctiondef}, we can establish the unifying relations between gluons and minimally coupled scalars \eqref{eq:unifyingminicouplescalarandYM}.


Our proof centers around the recursive BG currents, which we have previously argued, can be represented by a unique binary tree and can be related by differential operators. Specifically, the differential operators $\ma{T}^X$ in \eqref{eq:unifyingminicouplescalarandYM} can connect the BG currents for gluons and minimally coupled scalars, as shown below:
\begin{equation}\label{eq:unifyingcurrentYMandminScalar}
	\begin{split}
		\mathcal{T}^{In}\mathcal{A}_{I}\cdot\epsilon_{n}&=(-1)^{\frac{|I|-1}{2}}\phi_{I}\ \ (|I|\ \text{odd}),\\
		\mathcal{T}^{I}\mathcal{A}_{I}\cdot v&=(-1)^{\frac{|I|}{2}}\tilde{\mathcal{A}}_{I}\cdot v\ \ (|I|\ \text{even}).
	\end{split}
\end{equation}
Here, $\epsilon_n$ is the single particle polarization function for gluons, and $\phi_I$ is the wave function for scalars, both used as initial conditions. The tilde represents gluons minimally coupled to scalars. The vector $v_\mu$ can be any vector, such as a momentum vector or a polarization vector. Importantly, we can show that for each given deconcatenation, the relations \eqref{eq:unifyingcurrentYMandminScalar} still hold for the total BG currents, which are the sum of all possible deconcatenations. We can also generalize these relations to include partial deconcatenations, as shown below:
\begin{equation}\label{eq:unifyingcurrentYMandminScalarpartial}
	\begin{split}
		\mathcal{T}^{In}\mathcal{A}^{\Gamma}_{I}\cdot\epsilon_{n}&=(-1)^{\frac{|I|-1}{2}}\phi^{\Gamma}_{I}\ \ (|I|\ \text{odd}),\\
		\mathcal{T}^{I}\mathcal{A}^{\Gamma}_{I}\cdot v&=(-1)^{\frac{|I|}{2}}\tilde{\mathcal{A}}^{\Gamma}_{I}\cdot v\ \ (|I|\ \text{even}).
	\end{split}
\end{equation}
The superscript $\Gamma$ denotes a specific deconcatenation or binary tree. For instance, the 3-point BG current has two distinct deconcatenations: $\Gamma=[[1,2],3]$ and $\Gamma=[1,[2,3]]$. Hence, we only need to demonstrate that the BG current $\ma{A}_I^{\Gamma}$ satisfies \eqref{eq:unifyingcurrentYMandminScalarpartial} for both $\Gamma=[[1,2],3]$ and $\Gamma=[1,[2,3]]$. Then, the total BG currents $\ma{A}_I=\ma{A}_I^{[[1,2],3]}+\ma{A}_I^{[1,[2,3]]}$ would clearly satisfy \eqref{eq:unifyingcurrentYMandminScalar}. The same approach can be employed to higher point BG currents by induction. The calculations involved in the induction process are straightforward. For more information, readers can refer to our prior work \cite{Tao:2022nqc}. Once we establish the connections between BG currents \eqref{eq:unifyingcurrentYMandminScalar}, the proof for the unifying relation in the correlation function perspective \eqref{eq:unifyingminicouplescalarandYM} can also be carried out effortlessly. This is because we only need to substitute the first equation in \eqref{eq:unifyingcurrentYMandminScalar} into the correlation function definition \eqref{eq:correlationfunctiondef}.

In the forthcoming discussion, we will demonstrate that the unifying relation \eqref{eq:unifyingminicouplescalarandYM} can be extended to the loop level, and the proof is built upon the tree level outcome using the ``sewing procedure". This technique instructs us on how to construct the loop integrand from the tree level correlation function.
%

\section{Loops in Minkowski spacetime}\label{sec:flat}
We will begin by utilizing the off-shell currents at the tree level to construct loop integrands using the sewing procedure outlined in \cite{Gomez:2022dzk}. This section will provide an overview of the sewing procedure for BG currents in Minkowski spacetime. To begin, we select the recursive BG currents with all external legs off-shell. Next, we sew two legs together to obtain a loop integrand. Finally, we manually recover the cyclic completion and eliminate any unnecessary terms. We will use the example of the YM theory to demonstrate the sewing procedure in more detail.

The perturbiner methods for YM theory were initially proposed for multi-gluon scattering in \cite{Berends:1987me}. We employ these perturbiner methods to derive the BG currents for pure YM theory. However, there are some subtleties that need to be emphasized before we begin our calculation. While the gauge fixing is straightforward for the derivation of recursive BG currents at the tree level, it needs to be considered carefully at the loop level, taking into account the Faddeev-Popov ghosts. In contrast to Section \ref{sec:flatprooftree}, we consider the $\xi=1$ gauge in the following discussion and focus primarily on the gluon sector. For the ghost contribution, interested readers can refer to the previous literature \cite{Gomez:2022dzk}. The Lagrangian containing the gauge fixing term can be written as:
\begin{equation}
    \ma{L}=-\frac{1}{4}\tr \mb{F}_{\mu\nu}\mb{F}^{\mu\nu}-\frac{1}{2}(\partial_{\mu}\mb{A}^{\mu})^{2}+\partial^{\mu}b(\partial_{\mu}c-i[\mb{A}_{\nu},c]).
\end{equation}
In the Lagrangian, $\mb{F}_{\mu\nu}=\partial_{\mu}\mb{A}_{\nu}-\partial_{\nu}\mb{A}_{\mu}-i[\mb{A}_{\mu},\mb{A}_{\nu}]$ represents the field strength tensor, where $\mb{A}_{\mu}=A_{\mu}^aT^a$ is the Lie algebra valued gauge field and $b$ and $c$ are the Lie-algebra valued ghost fields. In order to derive the recursive BG currents, we first need to write down the equation of motion for the gauge fields in the $\xi=1$ gauge:
\begin{equation}\label{eq:gluoneomflat}
    \square \mb{A}_{\mu}=-i[\mb{A}^{\nu},\mb{F}_{\mu\nu}]+i\partial_{\nu}[\mb{A}^{\nu},\mb{A}_{\mu}]-i\{\partial_{\mu}b,c\}.
\end{equation}
The multi-particle solution to \eqref{eq:gluonansatznoghost} can be found using the same perturbiner expansion ansatz \eqref{eq:gluonansatznoghost} as in the Lorenz gauge. The ghost contributions can be calculated similarly to what was done for gauge fields, but we will not delve into the details here. By substituting the ansatz for both gauge and ghost fields \eqref{eq:gluonansatznoghost} into the equation of motion \eqref{eq:gluoneomflat}, we can obtain the off-shell currents for YM theory with ghost field contributions:
\begin{equation}
	\begin{split}
		s_P\ma{A}_P^{\mu}=&\sum_{P=QR}[k_R^{\mu}b_Rc_Q+\ma{A}_Q^{\nu}\ma{A}_R^{\rho}(k_{P\nu}\delta^{\mu}_{\rho}+k_{R\nu}\delta^{\mu}_{\rho}+k_Q^{\mu}\eta_{\nu\rho})-(Q\leftrightarrow R)]\\
		&+\sum_{P=QRS}[\ma{A}_Q^{\nu}\ma{A}_R^{\rho}\ma{A}_S^{\sigma}(\eta_{\nu\sigma}\delta^{\mu}_{\rho}-\eta_{\nu\rho}\delta^{\mu}_{\sigma})+(Q\leftrightarrow S)].
	\end{split}	
\end{equation}
In the above expression, the Mandelstam variables are denoted as $s_P=-k_P^2$, and the sums over $P=RS$ and $Q=TU$ represent the deconcatenation of the word $P$ into $R$ and $S$, and the word $Q$ into $T$ and $U$, respectively. To clarify this, let us take the example of a three-particle solution again where $P=ijk$. Then, $P=RS$ implies that we have $(R,\ S)=(i,\ jk)$ and $(ij,\ k)$ as possible deconcatenations of $P$.
%


To continue our discussion, let's make some remarks on the BG currents, which include both gauge fields and ghost fields. It is worth noting that the momentum $k$ will not be null for the off-shell gluon leg. However, it is important to emphasize that this current only coincides with the BG current at the multi-particle level. For single-particle states, the off-shell currents only yield a formal vector $\ma{A}_{i\mu}$, not the polarization vector $\epsilon_{i\mu}$. For simplicity, we will not distinguish between $\ma{A}_{i\mu}$ and $\epsilon_{i\mu}$ from now on. We just need to keep in mind that $k_{i}\cdot\epsilon_{i}$ vanishes for on-shell legs while it does not for off-shell legs. To construct the loop integrand, we need to first single out one leg and sew it together with another off-shell leg to generate the entire loop integrand. To illustrate this explicitly, we can rewrite the BG currents $P'=lP$ with a single gluon leg $l$:
\begin{equation}\label{eq:ymsewingminkowski}
	\begin{split}
		s_{P'}\ma{A}_{P'}^{\mu}=&[k_P^{\mu}b_Pc_l+\ma{A}_l^{\nu}\ma{A}_P^{\rho}(k_{lP\nu}\delta^{\mu}_{\rho}+k_{P\nu}\delta^{\mu}_{\rho}+k_l^{\mu}\eta_{\nu\rho})-(l\leftrightarrow P)]\\
		&+\sum_{P=QR}[\ma{A}_l^{\nu}\ma{A}_Q^{\rho}\ma{A}_R^{\sigma}(\eta_{\nu\sigma}\delta^{\mu}_{\rho}-\eta_{\nu\rho}\delta^{\mu}_{\sigma})+(l\leftrightarrow R)]\\
		&+\sum_{P=QR}[k_{lQ}^{\mu}b_{lQ}c_R+\ma{A}_{lQ}^{\nu}\ma{A}_R^{\rho}(k_{lQR\nu}\delta^{\mu}_{\rho}+k_{R\nu}\delta^{\mu}_{\rho}+k_{lQ}^{\mu}\eta_{\nu\rho})-({lQ}\leftrightarrow R)]\\
		&+\sum_{P=QRS}[\ma{A}_{lQ}^{\nu}\ma{A}_R^{\rho}\ma{A}_S^{\sigma}(\eta_{\nu\sigma}\delta^{\mu}_{\rho}-\eta_{\nu\rho}\delta^{\mu}_{\sigma})+({lQ}\leftrightarrow S)]
	\end{split}	
\end{equation}
We aim to consider the gluon loop, which means that we need to take $b_{l}=c_{l}=0$. Here we define $\ma{J}_{P\mu\nu} $ as $\ma{A}_{P'}^{\mu}=\epsilon_{l}^{\nu}\ma{J}_{P\mu\nu}$. Then we can obtain the recursion relation of $\ma{J}_{P\mu\nu}$:
%
\begin{equation}
	\begin{split}
		s_{P'}\ma{J}_{P\mu\nu}=&\ma{A}_{P\rho}[\delta^{\rho}_{\mu}(k_{lP}+k_P)_{\nu}+\delta^{\rho}_{\nu}(k_l-k_P)_{\mu}-\eta_{\mu\nu}(k_l+k_{lP})^{\rho}]\\
		&+\sum_{P=QR}(2\delta^{\rho}_{\mu}\delta^{\sigma}_{\nu}-\delta^{\sigma}_{\mu}\delta^{\rho}_{\nu}-\eta_{\mu\nu}\eta^{\rho\sigma})\ma{A}_{Q\rho}\ma{A}_{R\sigma}\\
		&+\sum_{P=QR}[\delta^{\sigma}_{\mu}(k_{lP}+k_R)^{\rho}-\delta^{\rho}_{\mu}(k_{lP}+k_{lQ})^{\sigma}+\eta^{\rho\sigma}(k_l+k_Q-k_R)_{\mu}]\ma{J}_{Q\rho\nu}\ma{A}_{R\sigma}\\
		&+(2\delta^{\rho}_{\gamma}\delta^{\sigma}_{\mu}-\eta^{\rho\sigma}\eta_{\mu\gamma}-\delta^{\sigma}_{\gamma}\delta^{\rho}_{\mu})\sum_{P=QRS}\ma{J}_{Q\rho\nu}\ma{A}_{R\sigma}\ma{A}^{\gamma}_S.
	\end{split}
\end{equation}


The object that we want to sew can be obtained from the BG currents by contracting the polarization vector of the off-shell leg $\epsilon_n^{\mu}$ and enforcing momentum conservation. This can be expressed mathematically as:
\begin{equation}
	A(l,P,n)=\lim\limits_{k_{lPn}\to0}s_{lP}\ma{A}_{lP\mu}\epsilon_n^{\mu}=\lim\limits_{k_{lPn}\to0}s_{lP}(\epsilon_l^{\nu}\ma{J}_{P\mu\nu})\epsilon_n^{\mu}.
\end{equation}
Here, $s_{lP}$ represents the Mandelstam variable, and $\ma{J}_{P\mu\nu}$ is called the ``one-loop pre-integrand" for one gluon loop. The limit $k_{lPn}\to 0$ implies the enforcement of momentum conservation.


The above analysis demonstrates that the sewing procedure can be executed in the following manner. First, we replace the polarization vectors of the two off-shell legs with the propagator, i.e., $\epsilon_l^{\mu}\epsilon_n^{\mu}\to\eta^{\mu\nu}/k_l^2$. Then, we enforce the momentum conservation condition, $k_l^{\mu}=-k_n^{\mu}$. This yields an integrand that resembles a one-loop diagram, expressed as $I^{\text{1-loop}}(P)\approx\eta^{\mu\nu}\ma{J}_{P\mu\nu}$ for a single trace color-ordered correlator. It should be noted that the first two terms in \eqref{eq:ymsewingminkowski} after the sewing contribute to the tadpole, and we must remove such diagrams explicitly. Finally, we can write the one-loop gluon integrand as:
\begin{equation}\label{eq:1loopinterandgluon}
	I_{\text{gluon}}^{\text{1-loop}}(P;l)=\eta^{\mu\nu}\tilde{\ma{J}}_{P\mu\nu}(l).
\end{equation}
We should note again that we have enforced the momentum conservation condition $k_P^{\mu}=0$, and $l$ denotes the loop momentum, i.e., $l^{\mu}=k_l^{\mu}$. The expression for $\tilde{\ma{J}}_{P\mu\nu}(l)$ is given by:
\begin{equation}
	\begin{split}
		\tilde{\ma{J}}_{P\mu\nu}(l)=&\frac{1}{l^2}\sum_{P=[QR]}\ma{J}_{Q\rho\mu}\ma{A}_{R\sigma}[\delta^{\sigma}_{\nu}(k_R+l)^{\rho}-\delta^{\rho}_{\nu}(k_Q+2l)^{\sigma}+\eta^{\rho\sigma}(2k_Q+l)_{\nu}]\\
		&+\frac{1}{l^2}\sum_{P=[QRS]}\ma{J}_{Q\rho\mu}\ma{A}_{R\sigma}\ma{A}_{S\gamma}(2\delta^{\sigma}_{\nu}\eta^{\gamma\rho}-\eta^{\sigma\gamma}\delta^{\rho}_{\nu}-\delta^{\gamma}_{\nu}\eta^{\rho\sigma})
	\end{split}
\end{equation}
%
Here, the notation $[QR]$ denotes the cyclic completion, which means we need to sum over all inequivalent cyclic permutations of a given deconcatenation of $P$. When $P=[RS]$, we first deconcatenate $P$ into the non-empty words $R$ and $S$, and then add $|S|-1$ cyclic permutations in $P$. For instance, if $P$ is the word $1234$, then the possible deconcatenations are:
\begin{equation}
	(R,S)=(1,234),(12,34),(123,4).
\end{equation}
On the other hand, the operation $P=[RS]$ gives:
\begin{equation}
	\begin{split}
		(R,S)=&(1,234),(2,134),(3,412),\cancel{(4,123)},\\
		&(12,34),(23,41),\cancel{(34,12)},\cancel{(41,23)},\\
		&(123,4),\cancel{(234,1)},\cancel{(341,2)},\cancel{(412,3)}.
	\end{split}
\end{equation}
The term with a backslash is redundant and can be expressed as one of the terms that remain. Additionally, when applying the operation $P=[QRS]$, we sum over all possible deconcatenations and add the cyclic permutation, which means we don't need to exclude any terms.

So far, we have only considered the contributions from the gluon sector. However, at the one-loop level, the ghost contributions cannot be ignored. It is worth mentioning that the ghost sewing procedure is quite similar to that of gluons. In the following discussion, we present the main results.

Let us begin by writing down the BG currents for the ghost fields, which can be approached using a multi-particle ansatz similar to the gluons. The ghost fields BG currents can be expressed as follows:
\begin{align}
	-s_{P}b_{P}&=\sum_{P=QR}b_{Q}(k_{Q}\cdot\ma{A}_{R})-(Q\leftrightarrow R)\\
	-s_{P}c_{P}&=\sum_{P=QR}c_{Q}(k_{P}\cdot\ma{A}_{R})-(Q\leftrightarrow R)
\end{align}
Following the sewing procedure, we proceed to identify an off-shell ghost leg $l$ as either $b_l$ or $c_l$, such that $b_{lP}=\ma{B}_P$ and $c_{lP}=c_l\ma{C}_P$. The expressions for $\ma{B}_P$ and $\ma{C}_P$ are as follows:
\begin{align}
	-s_{lP}\ma{B}_{P}&=k_{l}\cdot\ma{A}_{P}+\sum_{P=QR}\ma{B}_{Q}(k_{lQ}\cdot\ma{A}_{R})\\
	-s_{lP}\ma{C}_{P}&=k_{lP}\cdot\ma{A}_{P}+\sum_{P=QR}\ma{C}_{Q}(k_{P}\cdot\ma{A}_{R})
\end{align}
It is important to note that $\ma{B}_P$ and $\ma{C}_P$ only involve gluons, since the ghost polarization has been removed. Additionally, we ignore the contributions from the first term in both $\ma{B}_P$ and $\ma{C}_P$ as they only contribute to the tadpole. 

After obtaining the off-shell ghost leg $l$, the next step is to sew it by replacing its formal polarization vectors with the propagator. Then, we need to recover the cyclic completion manually and eliminate redundant terms. By doing so, we can construct the one-loop integrand for ghost fields:
\begin{align}
	\tilde{\ma{B}}_{P}&=-\frac{1}{l^2}\sum_{P=[QR]}\ma{B}_{Q}(k_{lQ}\cdot\ma{A}_{R}),\\
	\tilde{\ma{C}}_{P}&=-\frac{1}{l^2}\sum_{P=[QR]}\ma{C}_{Q}(k_{l}\cdot\ma{A}_{R}).
\end{align}
It should be noted that tadpole contributions have been ignored. Additionally, when the external gluon is taken to be on-shell, the two different ghost one-loop integrands coincide, i.e., $\tilde{\ma{B}}_{P}=\tilde{\ma{C}}_{P}$. Therefore, the total one-loop integrand for gluons can be written as:
\begin{equation}
	I^{\text{1-loop}}(P;l)=I_{\text{gluon}}^{\text{1-loop}}(P;l)-\tilde{\ma{C}}(P;l)
\end{equation}
The one-loop integrand for ghosts contributes to an additional minus sign, since the ghost loop behaves as a fermionic loop.

We should note that the gluon one-loop integrand \eqref{eq:1loopinterandgluon} contributes to the single-trace partial amplitudes. And the total one-loop amplitudes for gluons can be systematically constructed from Del Duca-Dixon-Maltoni color decomposition \cite{DelDuca:1999rs,Du:2014uua}. Let us consider the construction of total gluon amplitudes from its partial amplitudes $A_{n;0}$ with $n=|P|$:
\begin{equation}
    A_{\text{total}}^{\text{1-loop}}=\sum_{\sigma\in S_{n-1}/\ma{R}}c_n(\sigma)A_{n;0}(\sigma_1,\cdots,\sigma_n).
\end{equation}
Here $\sigma$ is the color order, and $c_n(\sigma)$ is the color basis defined by nested commutators of the group generators:
\begin{equation}
    c_n(\sigma)=\tr (T^a[T^{a_{\sigma_1}},[\cdots,[T^{a_{\sigma_{n-1}}},[T^{a_{\sigma_n}},T^a]]\cdots]]).
\end{equation}
And in the sum, $S_{n-1}$ represents the permutations of $(n-1)$ legs and $\ma{R}$ denotes the reflection. The partial amplitude $A_{n;0}$ is what we consider in this paper.


\textbf{Summary:} To summarize the sewing procedure in Minkowski spacetime, we first compute the BG currents with all external legs off-shell, using a specific equation of motion. Then, we single out an external leg from the word of the BG currents, usually the first letter, and take the other legs on-shell, leaving 2 off-shell legs. Next, we sew the two off-shell legs by replacing their formal polarization vectors with the propagator. It is essential to recover the cyclic completion by hand and remove redundant terms. However, there is an ambiguity that requires careful consideration. It is best to carry out all calculations in the $\xi=1$ gauge instead of the Lorenz gauge. This is because in the second step, we select a leg to be off-shell, and the Lorenz gauge can cause confusion due to the transverse condition. Therefore, choosing the $\xi=1$ gauge is convenient to emphasize that we have no transverse condition.

\section{Differential Operators and Unifying relations at the 1-loop Level in Flat Spacetime}\label{sec:flatproof}
In this section, we will show the unifying relations at the 1-loop level using the sewing procedure on BG currents in flat spacetime. The unifying relations are valid at the BG currents level, which suggests that they can be generalized to the loop integrand level as the latter can be constructed from the BG currents. In the following discussion, we will consider specific theories and prove the one-loop unifying relation by recursion. 

\subsection{Minimal coupled scalar}
Let us begin with a straightforward case. Consider the scalar theory with minimal coupling with gluons. In this case, we can obtain the off-shell scalar currents in the $\xi=1$ gauge:
\ie\label{eq:bgcurrentscalarymminkowski}
s_{P}\phi_{P}=&\sum_{P=QR}[(k_{P}\cdot \tilde{\ma{A}}_{Q}\phi_{R}+k_{R}\cdot \tilde{\ma{A}}_{Q}\phi_{R})-(Q\leftrightarrow R)]\\
&+\sum_{P=QRS}[\tilde{\ma{A}}_{Q}\cdot \tilde{\ma{A}}_{S}\phi_{R}-\tilde{\ma{A}}_{Q}\cdot \tilde{\ma{A}}_{R}\phi_{S}+(Q\leftrightarrow S)],
\fe
where the gluon currents with a tilde denote the gluon currents, including scalar coupling terms, i.e.,
\ie\label{eq:gluonbgcurrentmodified}
s_P\tilde{\ma{A}}_{P}^{\mu}=&\sum_{P=QR}[k_R^{\mu}b_Rc_Q+\tilde{\ma{A}}_Q^{\nu}\tilde{\ma{A}}_R^{\rho}(k_{P\nu}\delta^{\mu}_{\rho}+k_{R\nu}\delta^{\mu}_{\rho}+k_Q^{\mu}\eta_{\nu\rho})+k_{R}^{\mu}\phi_{R}\phi_{Q}-(Q\leftrightarrow R)]\\
&+\sum_{P=QRS}[\tilde{\ma{A}}_Q^{\nu}\tilde{\ma{A}}_R^{\rho}\tilde{\ma{A}}_S^{\sigma}(\eta_{\nu\sigma}\delta^{\mu}_{\rho}-\eta_{\nu\rho}\delta^{\mu}_{\sigma})+\tilde{\ma{A}}_{Q}^{\mu}\phi_{R}\phi_{S}-\tilde{\ma{A}}_{R}^{\mu}\phi_{Q}\phi_{S}+(Q\leftrightarrow S)].
\fe
To recap the sewing procedure, once we have the BG current \eqref{eq:bgcurrentscalarymminkowski}, the next step is to select an external leg from the word of the BG current \eqref{eq:bgcurrentscalarymminkowski}. We can choose a leg $l$ to be the only scalar leg and sew it with leg $n$, the only off-shell leg in the traditional BG currents, in the subsequent step by setting $P' = lP$ (all legs in $P$ are gluons). Here, we write down the current with an explicit label $l$:
\ie
s_{P'}\phi_{P'}=&(k_{lP}\cdot \tilde{\ma{A}}_{l}\phi_{P}+k_{P}\cdot \tilde{\ma{A}}_{l}\phi_{P})-(l\leftrightarrow P)\\
&+\sum_{P=QR}[\tilde{\ma{A}}_{l}\cdot \tilde{\ma{A}}_{R}\phi_{Q}-\tilde{\ma{A}}_{l}\cdot \tilde{\ma{A}}_{Q}\phi_{R}+(l\leftrightarrow R)]\\
&+\sum_{P=QR}[(k_{lP}\cdot \tilde{\ma{A}}_{lQ}\phi_{R}+k_{R}\cdot \tilde{\ma{A}}_{lQ}\phi_{R})-(lQ\leftrightarrow R)]\\
&+\sum_{P=QRS}[\tilde{\ma{A}}_{lQ}\cdot \tilde{\ma{A}}_{S}\phi_{R}-\tilde{\ma{A}}_{lQ}\cdot \tilde{\ma{A}}_{R}\phi_{S}+(lQ\leftrightarrow S)]
\fe
Next, we put the particles in the word $P$ on-shell and let $\phi_{l}\Phi_{P}=\phi_{lP}$. For the purpose of this intuitive discussion, we only consider a loop with a scalar particle running in it. Therefore, we assume that $\tilde{\ma{A}}_{lP}$ vanishes; otherwise, the resulting loop would include gluon propagators. We then have:
\ie
s_{lP}\Phi^{s}_{P}=&-k_{lP}\cdot \tilde{\ma{A}}_{P}-k_{l}\cdot \tilde{\ma{A}}_{P}-\sum_{P=QR}(\tilde{\ma{A}}_{Q}\cdot \tilde{\ma{A}}_{R})\\
&+\sum_{P=QR}[-k_{lP}\cdot \tilde{\ma{A}}_{R}\Phi^{s}_{Q}-k_{1Q}\cdot \tilde{\ma{A}}_{R}\Phi^{s}_{Q}]+\sum_{P=QRS}[-\tilde{\ma{A}}_{R}\cdot \tilde{\ma{A}}_{S}\Phi^{s}_{Q}].
\fe
The superscript $s$ in the expression above indicates that we are only considering scalar loop contributions. Following the procedure outlined in Section \ref{sec:flat}, we can express the pre-integrand of the loop diagram (after stripping the tadpole terms and setting $k_{P}^{\mu}=0$) as:
\ie
-\tilde{\Phi}^{s}_{P}=\frac{1}{l^2}\{\sum_{P=[QR]}(k_{Q}+2l)\cdot \tilde{\ma{A}}_{R}\Phi^{s}_{Q}+\sum_{P=[QRS]}\tilde{\ma{A}}_{R}\cdot \tilde{\ma{A}}_{S}\Phi^{s}_{Q}\}.
\fe
The modified BG current for gluons, taking into account their interaction with scalars, is denoted by $\tilde{\ma{A}}_R$ as given by \eqref{eq:gluonbgcurrentmodified}. It is worth noting that the unifying relation between pure YM and the theory of scalars with a minimal coupling with gluons is valid at the tree level even for their BG currents. Furthermore, since the one-loop integrand can be constructed from the BG currents, we can expect that some relations can be inherited from the BG currents after applying the sewing procedure. To illustrate these potential relations, let us reiterate the one-loop integrand for pure gluons:
\begin{equation}
	\begin{split}
		\eta^{\mu\nu}\tilde{\ma{J}}_{P\mu\nu}=&\frac{1}{l^2}\sum_{P=[QR]}\ma{J}_{Q\rho\sigma}\ma{A}_{R\sigma}(k_R+l)^{\rho}-\ma{J}_{Q\rho}^{\rho}\ma{A}_{R\sigma}(k_Q+2l)^{\sigma}+\ma{J}_{Q}^{\rho\nu}\ma{A}_{R\rho}(2k_Q+l)_{\nu}]\\
		&+\frac{1}{l^2}\sum_{P=[QRS]}2\ma{J}_{Q\rho}^{\sigma}\ma{A}_{R\sigma}\ma{A}_{S}^{\rho}-\ma{J}_{Q\rho}^{\rho}(\ma{A}_{R}\cdot\ma{A}_{S})-\ma{J}_{Q\rho\sigma}\ma{A}_{R}^{\rho}\ma{A}_{S}^{\sigma}
	\end{split}
\end{equation}

Let us consider the following operator:
\ie
	\ma{D}=\frac{\partial}{\partial d},
\fe
where $d$ is the dimension that arises from $\eta_{\mu\nu}\eta^{\mu\nu}$. The terms that depend on $d$ can only be found in $\ma{J}_{\mu\nu}\eta^{\mu\nu}$. Unifying relations for BG currents also hold for $\xi=1$ gauge using the method in \cite{Tao:2022nqc}. Thus, if we wish to transform the gluon loop into the scalar loop, we need to apply $\ma{T}[ln]$ to the pure gluon currents first to ensure that the legs $l$ and $n$ are scalars in the same trace, meaning there is no gluon propagator between them, and then sew these two legs together. The only surviving terms in $\ma{A}_{lP}\cdot\epsilon_{n}=\epsilon_{l\mu}\epsilon_{n\nu}\ma{J}_{P}^{\mu\nu}$ after applying $\ma{T}[ln]$ correspond to the terms in $\ma{J}_{P\mu\nu}$ that are proportional to $\eta_{\mu\nu}$. Therefore, if we replace $\epsilon_{l\mu}\epsilon_{n\nu}$ with $\eta_{\mu\nu}/l^2$, we obtain terms proportional to $\eta_{\mu\nu}\eta^{\mu\nu}=d$. Thus, the operator $\ma{D}$ acting on the loop amplitudes is equivalent to the operator $\ma{T}[ln]$ acting on the BG currents and then sewing the legs $l$ and $n$ together. Based on the analysis above, it is evident that only the terms that are proportional to $\ma{J}_{P\rho}^{\rho}$, which inherently contain $\eta^{\rho}_{\rho}=d$, will survive. Therefore, the surviving terms can be expressed as:
%
\begin{equation}
	\begin{split}
		\eta^{\mu\nu}\tilde{\ma{J}}_{P\mu\nu}\sim&\frac{1}{l^2}\sum_{P=[QR]}-\ma{J}_{Q\rho}^{\rho}\ma{A}_{R\sigma}(k_Q+2l)^{\sigma}+\frac{1}{l^2}\sum_{P=[QRS]}-\ma{J}_{Q\rho}^{\rho}(\ma{A}_{R}\cdot\ma{A}_{S})
	\end{split}.
\end{equation}

After examining the observation above, it appears that we can propose the following ansatz:
\ie\label{uni}
		\ma{D}\ma{J}_{P\rho}^{\rho}=\Phi^{s}_{P}.
\fe
The ansatz proposed above involves the scalar loop integrand $\Phi$ with all external legs being gluons. It can be easily proven by induction, providing us with a method to transmute a gluon loop to a scalar loop. However, the task becomes more complicated if we aim to transmute the gluon legs to scalar legs as well. At the tree level, the unifying relation between the Yang-Mills theory and the scalar theory with minimal gluon coupling involves a sum of operators. The sewing procedure, however, requires us to take a trace operator $\ma{T}[ln]$ if we want to obtain a scalar loop. This means that we cannot obtain all the operators in the sum, and the unifying relation fails. Therefore, if we wish to transmute both the loop and the legs, we cannot have the scalar running in the loop and the external scalars in the same trace. In summary, this can be expressed as follows:
\ie
		\ma{T}^{X}\ma{D}\ma{S}=\ma{S}\ma{T}^{X}\ma{T}[ln]
\fe
In the equation above, $\ma{S}$ represents the sewing procedure. This implies that the unifying relation still holds when we aim to obtain a loop integrand with a scalar loop and any combination of external legs. In the subsequent discussions, we will concentrate on this particular scenario.

Before concluding this section, there is an important question that needs to be addressed: whether the ghost loop contributes to the unifying relation operators. In fact, when we consider the total YM loop-integrand, we must take into account ghost loops. Thus, the behavior of the ghost loops after acting $\ma{D}$ is important for the unifying relation. We must pay attention to this possible contribution. To illustrate this subtlety, let us explicitly write down again the one-loop integrand for all the ghost fields:
\begin{align}
	\tilde{\ma{B}}_{P}&=-\frac{1}{l^2}\sum_{P=[QR]}\ma{B}_{Q}(k_{lQ}\cdot\ma{A}_{R}),\\
	\tilde{\ma{C}}_{P}&=-\frac{1}{l^2}\sum_{P=[QR]}\ma{C}_{Q}(k_{l}\cdot\ma{A}_{R}).
\end{align}
To clarify this point, we should keep in mind that the operator $\ma{D}$ acts only on terms proportional to $\eta_{\mu\nu}\eta^{\mu\nu}$. However, the ghost one-loop integrand does not contain any terms that could provide such a contribution. Therefore, we can safely conclude that the ghost one-loop integrand does not contribute to the unifying relations, although it does contribute to the total one-loop integrand. It is important to keep this in mind when considering the behavior of ghost loops.

We need to provide further information on the tadpole. In Minkowski space, the massless particle tadpole cancels out after the dimensional regularization. However, in curved spacetime, the curvature introduces an effective mass term to the massless particle, leading to a non-vanishing tadpole contribution. Despite this complication, we can explain the tadpole contribution simply by writing down the one-loop integrand attributed to the tadpole and determining its impact on the unifying relations. For the YM theory, we can express the tadpole as follows:
\ie
\ma{A}_{P\rho}[(k_{lP}+k_P)^{\rho}+(k_l-k_P)^{\rho}-d(k_l+k_{lP})^{\rho}]+\sum_{P=QR}(1-d)\ma{A}_{Q}\cdot\ma{A}_{R}.
\fe
The integrand for the one-loop tadpole contribution in the scalar theory can be expressed as:
\ie
-k_{lP}\cdot \tilde{\ma{A}}_{P}-k_{l}\cdot \tilde{\ma{A}}_{P}-\sum_{P=QR}(\tilde{\ma{A}}_{Q}\cdot \tilde{\ma{A}}_{R}).
\fe
This demonstrates that the two tadpole contributions are exactly the same after the application of the unifying operator $\ma{D}$, implying that the tadpole does not affect our previous argument. Therefore, for our objective of constructing the differential operators at the one-loop integrand level, we can safely disregard the tadpole contribution in our subsequent discussions.


\subsection{Yang-Mills scalar theory}
In fact, the situation is more favorable for the general YMS theory, as we do not need to sum over operators and can easily transform both loops and legs. In this subsection, we will construct the BG currents of the YMS theory first, and then show the sketch of the proof of the unifying relation for the YMS BG currents at the tree level. Based on this, we will then show the 1-loop unifying relation for the YMS theory. It's worth noting that we will not take into account the contributions from ghosts and tadpoles in the following discussion, as we have already discussed in the previous subsection. Let us now consider this more non-trivial case. The Lagrangian of the YMS theory is:
\ie
\ma{L}_{\text{YMS}}=&-\frac{1}{4}F^{a}_{\mu\nu}F^{a\mu\nu}+\frac{1}{2}(D_{\mu}\phi^{A})^{a}(D^{\mu}\phi^{A})^{a}\\
&-\frac{g^{2}}{4}f^{abe}f^{ecd}\phi^{Aa}\phi^{Bb}\phi^{Ac}\phi^{Bd}+\frac{1}{3!}\lambda gF^{ABC}f^{abc}\phi^{Aa}\phi^{Bb}\phi^{Cc}.
\fe
The Lagrangian of the YMS theory includes the gluon field strength $F^a_{\mu\nu}=\partial_{\mu}A^a_{\nu}-\partial_{\nu}A^a_{\mu}+gf^{abc}A^b_{\mu}A^c_{\nu}$, the covariant derivative $D_{\mu}\phi^{Aa}=\partial_{\mu}\phi^{Aa}+g f^{abc} A^b_{\mu}\phi^{Ac}$, and a scalar field $\phi^{Aa}$ that is charged under two gauge groups. While one of these groups is the familiar gauge group for gluons, the other group is of interest here. The generators for this group define a new Lie-algebra via the commutator:
\begin{equation}
	[\tilde{T}^A,\tilde{T}^B]=iF^{ABC}\tilde{T}^C,
\end{equation}
where the capital Latin index represents the generator and structure constant for the YMS gauge group. The gauge coupling for gluons is denoted by $g$, while the gauge coupling for YMS is denoted by $\lambda$. As we have argued that the second gauge group is tied to $U(1)$, in which $\tilde{T}^A\to1$, the YMS Lagrangian reproduces the one for gluons coupled to scalars minimally, after ignoring the higher-order self-interactions between scalars.

The equation of motion for the YMS scalars in the $\xi=1$ gauge, with the simplification of setting the YM coupling constant to $g=1$, is:
\ie
\square\phi=[\phi^{B},[\phi,\phi^{B}]]-\frac{1}{2}\lambda[[\phi,\phi]]-2i[\partial_{\mu}\phi,A^{\mu}]+i[\partial_{\mu}A^{\mu},\phi]-[[A^{\mu},\phi],A_{\mu}].
\fe
In the equation of motion, we consider the Lie-algebra valued scalar field $\phi$ with adjoint indices $A$ and $a$ associated with two different quadratic Lie algebras. Here, $\phi=\phi_{Aa}\tilde{T}^A\otimes T^a$. We also use the double bracket notation $[[\cdots]]$, which is defined as $[[T^a\otimes \tilde{T}^A,T^b\otimes \tilde{T}^B]]=[T^a,T^b]\otimes[\tilde{T}^A,\tilde{T}^B]$.

We can obtain the BG currents for YMS by considering the following perturbiner ansatz:
\ie
\phi&=\sum_{P}\phi_{P}T^{P}e^{ik_P\cdot x}\\
\tilde{A}_{\mu}&=\sum_{P}\tilde{A}_{P\mu}T^{P}e^{ik_P\cdot x}
\fe
and
\ie
&\phi_{P}=\sum_{\text{all\ possible\ traces }}\phi_{P|Q_{1},Q_{2},\cdots, Q_{m-1},Q_{m}}\prod_{i=1}^{m-1}\tr(\tilde{T}^{Q_{i}})\tilde{T}^{Q_{m}}\\
&\tilde{A}_{P\mu}=\sum_{\text{all\ possible\ traces}}\tilde{A}_{P|Q_{1},Q_{2},\cdots, Q_{m-1},Q_{m}\mu}\prod_{i=1}^{m-1}\tr(\tilde{T}^{Q_{i}}).
\fe
The perturbiner ansatz for obtaining the BG currents for YMS involves using the color factors $T^{P}=T^{p_{1}p_{2}\cdots p_{n}}$ and $\tilde{T}^{P}$ for the first and second colors of bi-adjoint scalars, respectively, as well as $Q_{m}$ for scalar currents that can include only one letter while other $Q_{i}$ require at least two letters. The word $Q$ can have a trace $\tr(\tilde{T}^{Q})$, which may be empty in which case we set it to 1. It is important to note that the coefficients or BG currents for the words in traces, such as $\tr(\tilde{T}^{Q})$, are invariant under cyclic permutation of $Q$. Thus, the sum of all possible traces is up to a cyclic permutation of the words in the traces, such as $\tr(T^{Q})$. This ansatz is used because the trace structure is determined by the Feynman rules and depends on the initial conditions imposed. When scalar external legs are chosen as initial conditions, their trace structures must be included as part of the initial conditions. Since particles in $Q_{m}$ are in the same trace as the off-shell leg $n$, generators in $Q_{m}$ do not need to be taken into the trace formalism. The trace part will take the trace operation only after multiplying the one-particle state of the leg $n$ and generating the amplitudes. Keeping the color factors after the trace operation yields the correct color-ordered amplitudes with the corresponding traces of the generators. The bi-adjoint property of the scalars allows for a more precise classification of the terms in the ansatz.

We choose the basis of the Lie algebra satisfying $\tr(T^{A}T^{B})=\delta^{AB}$, then we have
\ie
\phi_{P}^{B}\phi_{R}^{B}=\sum_{\text{all\ possible\ trace\ in}\ Q,\ S}&\phi_{P|Q_{1},Q_{2},\cdots, Q_{m-1},Q_{m}}\phi_{R|S_{1},S_{2},\cdots, S_{k-1},S_{k}}\\
&\times\prod_{i=1}^{m-1}\tr(\tilde{T}^{Q_{i}})\prod_{j=1}^{k-1}\tr(\tilde{T}^{S_{j}})\tr(\tilde{T}^{Q_{m}}\tilde{T}^{S_{k}})
\fe
We can write down the BG current:
\ie\label{yms}
-k_{P}^{2}\phi_{P|P_{1},P_{2},\cdots, P_{m-1},P_{m}}=&\sum_{P=XY}[\sum_{i=1}^{m-1}\sum_{P_{i}=JK}[\sum_{Y=CD}\phi_{C|P\backslash P_{i}\cap C,P_{m}}(\phi_{X|P\backslash P_{i}\cap X,J}\phi_{D|P\backslash P_{i}\cap D,K}\\
&~~~~~~+\sum_{KJ=RS}\phi_{X|P\backslash P_{i}\cap X,R}\phi_{D|P\backslash P_{i}\cap D,S})-(C\leftrightarrow D)]\\
&~~~~~~-\sum_{P_{m}=JK}\lambda(\phi_{X|P\cap X,J}\phi_{Y|P\cap Y,K})\\
&~~~~~~+\phi_{X|P\cap X,P_{m}}(k_{X}\cdot \tilde{A}_{Y|P\cap Y})+\phi_{X|P\cap X,P_{m}}(k_{P}\cdot \tilde{A}_{Y|P\cap Y})\\
&~~~~~~+\tilde{A}_{X|P\cap X}\cdot(\sum_{Y=CD}\tilde{A}_{C|P\cap C}\phi_{D|P\cap D,P_{m}}-(C\leftrightarrow D))]\\
&~~~~~~-[X\leftrightarrow Y]
\fe
The notation $P\cap X$ represents the traces in $P$ that contain only letters from the word $X$, and the notation $P\backslash P_i$ means that we remove the letters of $P_i$ from $P$. The term with $\sum_{KJ=RS}$ may be confusing, but it arises from terms like $\phi_X^A \phi_Y^A$ that already have a trace factor. Using $\tr(AB)=\tr(BA)$, we consider all possible combinations, leading to the sum $\sum_{KJ=RS}$. After this operation, the currents are invariant under the cyclic permutation of $P_i$. However, for the trace $P_m$ to which the off-shell leg belongs, we cannot do this operation because the generator of the off-shell leg is not included, and there is no trace operation for the series of the generator $T^{P_m}$. For gluon currents in the YMS theory, we also obtain
%
\ie
-k_{P}^{2}\tilde{A}_{P|P_{1},P_{2},\cdots, P_{m-1},P_{m}\mu}&=\sum_{P=XY}k_{Y\mu}(\tilde{A}_{X|P\cap X}\cdot \tilde{A}_{Y|P\cap Y})-\tilde{A}_{Y|P\cap Y\mu}(k_{Y}\cdot \tilde{A}_{X|P\cap X})\\
&-\tilde{A}_{Y|P\cap Y\mu}(k_{P}\cdot \tilde{A}_{X|P\cap X})-k_{Y\mu}\sum_{i=1}^{m}\sum_{P_{i}=JK}(\phi_{X|P\backslash P_{i}\cap X,J}\phi_{Y|P\backslash P_{i}\cap Y,K}\\
&+\sum_{KJ=RS}\phi_{X|P\backslash P_{i}\cap X,R}\phi_{Y|P\backslash P_{i}\cap Y,S})\\
&-\sum_{Y=CD}[\tilde{A}_{C|P\cap C\mu}[\tilde{A}_{X|P\cap X}\cdot \tilde{A}_{D|P\cap D}\\
&-\sum_{i=1}^{m}\sum_{P_{i}=JK}(\phi_{X|P\backslash P_{i}\cap X,J}\phi_{D|P\backslash P_{i}\cap D,K}+\sum_{KJ=RS}\phi_{X|P\backslash P_{i}\cap X,R}\phi_{D|P\backslash P_{i}\cap D,S})]\\
&-\tilde{A}_{D|P\cap D\mu}[\tilde{A}_{X|P\cap X}\cdot \tilde{A}_{C|P\cap C}-\sum_{i=1}^{m}\sum_{P_{i}=JK}(\phi_{X|P\backslash P_{i}\cap X,J}\phi_{C|P\backslash P_{i}\cap C,K}\\
&+\sum_{KJ=RS}\phi_{X|P\backslash P_{i}\cap X,R}\phi_{C|P\backslash P_{i}\cap C,S})]]-[X\leftrightarrow Y]
\fe

The relations between different scattering amplitudes can be demonstrated by establishing the corresponding relations between the BG currents \cite{Tao:2022nqc}. Specifically, if we set $\lambda=2$, the relations become more apparent. The specific coupling constant can be restored by taking into account the types of external legs and the trace structures of the currents. In fact, we have:
%
\ie\label{ymsuni}
		\mathcal{T}[P_{1}]\mathcal{T}[P_{2}]\cdots\mathcal{T}[P_{m}n](A_{P}\cdot\epsilon_{n})&=(-1)^{\sum_{i=1}^{m}|P_{i}|-m}\phi_{P|P_{1},\cdots,P_{m}},\\
		\mathcal{T}[P_{1}]\mathcal{T}[P_{2}]\cdots\mathcal{T}[P_{m}](v\cdot A_{P})&=(-1)^{\sum_{i=1}^{m}|P_{i}|-m}v\cdot \tilde{A}_{P|P_{1},\cdots,P_{m}}.
\fe
To distinguish the gluon currents in YMS theory and YM theory, we use the tilde notation. The factor $(-1)^{\sum_{i=1}^{m}|P_{i}|-m}$ arises from the opposite sign between some 3-point currents and the Feynman rules, including the $\phi^{3}$ vertices. Proving \eqref{ymsuni} is straightforward, and we highlight some essential points. Let us consider a sub-current $A_{X\mu}$ of the total BG current $A_{P\mu}$. If $P_{i}\subset X$, we have $\ma{T}[P_{i}]A_{X}\cdot v=(-1)^{|P_{i}|-1}A_{X|P_{i}}\cdot v$ and $\ma{T}[P_{i}n]A_{X}\cdot\epsilon_{n}=(-1)^{|P_{i}|-1}\phi_{X|P_{i}}$. If $P_{i}\cap X\neq \varnothing$ and $P_{i}\backslash X$ is in the word of another sub-current $A_{Y\mu}$, which is in the same term as $A_{X\mu}$, then we get a deconcatenation of $P_{i}=JK$, where $J\subset X$ and $K\subset Y$. If we cannot perform this deconcatenation by cyclic permutation, it leads to zero. If we can, this corresponds to the terms with $\sum_{KJ=RS}$ in the YMS currents. After applying the corresponding operators, we obtain $\ma{A}_{X|J\mu}\ (\phi_{X|J})$ and $\ma{A}_{Y|K\mu}\ (\phi_{Y|K})$. If $J$ is not a subset of $X$, then $\ma{A}_{X|J\mu}$ is zero, so a deconcatenation sum $\sum_{P_{i}=JK}$ is equivalent to selecting suitable $J$ and $K$ such that $J\subset X$ and $K\subset Y$, leading to the formalism in the YMS currents. Another important point is that when $\ma{T}[P_{i}]$ ($P_{i}\cap X\neq\varnothing$ and $P_{i}\cap Y\neq\varnothing$) acts on the term $A_{Y\mu}(k_{Y}\cdot A_{X})$, the word $P_{i}$ breaks into two pieces, and each current obtains a part. One can verify this by examining the operator $\ma{T}[P_{i}]$'s concrete form. With this process, we can prove \eqref{ymsuni}, similar to the process in \cite{Tao:2022nqc}.


Next, we proceed to sew a scalar loop. Without loss of generality, we choose the first letter in $P$ to be the scalar that will be sewn with the off-shell leg. This procedure can also be achieved by acting $\ma{T}[lP_{m}n]$ on the YM currents and then sewing the leg $l$ and the off-shell leg. Since there is a $\ma{T}[ln]$ in $\ma{T}[lP_{n}n]$, the operator for the loop integrand can be expressed as follows:
\ie\label{eq:ymsunifyingloop}
		\mathcal{T}[P_{1}]\mathcal{T}[P_{2}]\cdots\ma{T}[P_{m-1}]\partial_{k_{l}\cdot\epsilon_{\alpha_{1}}}\prod_{i=1}^{|P_{m}|-1}\partial_{k_{\alpha_{i}}\cdot\epsilon_{\alpha_{i+1}}}\ma{D}\ma{S}=\ma{S}\mathcal{T}[P_{1}]\mathcal{T}[P_{2}]\cdots\ma{T}[lP_{m}n],
\fe
%
where $\alpha_i$ denotes the $i$-th letter in $P_m$. It is worth noting that in BG currents, the momentum of the off-shell leg $k_{n\mu}$ does not appear explicitly. Hence, the operator $\mathcal{T}_{ijn}=\partial_{k_{i}\epsilon_{j}}-\partial_{k_{n}\epsilon_{j}}$ can also be written as $\partial_{k_{i}\epsilon_{j}}$ in the operator $\ma{T}[lP_{m}n]$.


In the subsequent analysis, our attention will be restricted to the conformally coupled scalar that is minimally coupled to gluons. This is a direct generalization of the particular instance of the YMS theory mentioned earlier, where we take the second color group of the scalars to be the $U(1)$ group.

\section{Loops in (A)dS spacetime}\label{sec:ads}

In this section, we examine the construction of loop integrands using the sewing procedure in (A)dS. As in Section \ref{sec:flat}, we first review the gluon equation of motion and BG currents in (A)dS. We anticipate that the sewing procedure can be generalized to curved spacetime based on our experience with it in flat space. We demonstrate the expected sewing procedure in (A)dS and provide concrete steps to sew a loop integrand from the tree-level currents through several examples. We then calculate some (A)dS gluon loop integrands using traditional Feynman rules and compare the results with those obtained from the sewing procedure. By sewing loop integrands from the tree-level BG currents, we expect to inherit the unifying relation from the tree-level correlation function \cite{Tao:2022nqc}. Finally, we consider gluons conformally coupled to scalars as an example to demonstrate the unifying relation at the loop level in (A)dS.

\subsection{BG currents in (A)dS}

In this subsection, we review the BG currents in (A)dS, following \cite{Armstrong:2022jsa}. We will consider $\text{AdS}_{d+1}$ in the Poincar\'{e} patch, where the metric can be expressed as follows:
\begin{equation}\label{eq:adsmetric}
	g_{mn}dx^mdx^n=\frac{\ma{R}^2}{z^2}(dz^2+\eta_{\mu\nu}dx^{\mu}dx^{\nu}),
\end{equation}
where the indices $m, n$ generally represent the radial direction $z$ and the boundary directions. The parameter $\ma{R}$ represents the radius, and the range of $z$ is $0<z<\infty$. The flat boundary metric $\eta_{\mu\nu}$ has a Lorentzian signature, and $\mu,\nu=0,1,\cdots,d-1$. It is worth noting that the $\text{dS}_{d+1}$ metric can be obtained from \eqref{eq:adsmetric} by analytically continuing $z\to -i\eta$ and $\ma{R}\to -i\ma{R}$ after taking the boundary metric to be Euclidean.

To maintain consistency with the flat case, we adopt the boundary gauge $\xi=1$ instead of the Lorenz gauge for the (A)dS case. Consequently, the equation of motion for gluons in the Poincar\'{e} patch can be expressed as follows:
\begin{equation}
    \begin{split}
        g^{np}\partial_p\mb{F}_{mn}=ig^{np}[\mb{A}_p,\mb{F}_{mn}]+\mb{J}_m+g^{np}\left(\Gamma^q_{mp}\mb{F}_{qn}+\Gamma^q_{np}\mb{F}_{mq}\right)-\frac{1}{\xi}g^{\mu\nu}\partial_{\rho}\delta^{\rho}_m\partial_{\mu}\mb{A}_{\nu}.
    \end{split}
\end{equation}
It should be noted that the index $\rho$ in the last term is not summed over and is simply a notation indicating that this term only contributes to the equation of motion for boundary terms. The last term represents the gauge fixing term. Here, $\mb{J}_m$ is the interaction current and $\Gamma^p_{mn}=g^{pq}\Gamma_{mnq}$ is the Christoffel symbol: 
\begin{equation}
	\Gamma_{mnp}=\frac{1}{2}\left(\partial_mg_{np}+\partial_n g_{mp}-\partial_p g_{mn}\right).
\end{equation} 
As we have mentioned in the Minkowski spacetime case, we also adopt the boundary $\xi=1$ gauge instead of the boundary transverse gauge to avoid ambiguities in the loop integrand sewing procedure.


We can apply the same perturbiner expansion ansatz as in the boundary Lorenz gauge to evaluate the multi-particle solution of the equation of motion in the $\xi=1$ gauge:
\begin{eqnarray}
	\mb{A}_{\mu}(x,z)&=&\frac{\ma{R}}{z}\sum_I \ma{A}_{I\mu}(z)T^{a_I}e^{ik_I\cdot x},\\
	\mb{A}_z(x,z)&=&\frac{\ma{R}}{z}\sum_I \alpha_I(z)T^{a_I}e^{ik_I\cdot x},\\
	\mb{J}_m(x,z)&=&\sum_I \mathcal{J}_{Im}(z)T^{a_I}e^{ik_I\cdot x}.
\end{eqnarray}
In this expression, the subscript $\mu$ denotes the boundary component and $z$ denotes the bulk component. By substituting the ansatz into the equation of motion, we obtain the recursion relation for gluons in the $\xi=1$ gauge:
\begin{equation}\label{eq:bgcurrentboundaryads}
	\begin{split}
		\frac{1}{z^2}(\ma{D}_I^2+d-1)\ma{A}_{I\mu}&=ik_{I\mu}(\partial_z+\frac{2-d}{z})\alpha_I-\frac{\ma{R}}{z}\mathcal{J}_{I\mu}+\frac{\ma{R}}{z}\sum_{I=JK}\{(k_{K\mu}\alpha_K+2i\partial_z\ma{A}_{K\mu})\alpha_J\\
		&+k_{K\mu}(\ma{A}_J\cdot\ma{A}_K)+\ma{A}_{K\mu}[i(\partial_z-\frac{d}{z})\alpha_J-k_I\cdot\ma{A}_J-k_K\cdot\ma{A}_J]-(J\leftrightarrow K)\}\\
		&+\frac{\ma{R}^2}{z^2}\sum_{I=JKL}\{[\alpha_J\alpha_K\ma{A}_{L\mu}+(\ma{A}_J\cdot\ma{A}_K)\ma{A}_{L\mu}-(K\leftrightarrow L)]\\
		&+[\alpha_K\alpha_L\ma{A}_{J\mu}+(\ma{A}_K\cdot\ma{A}_L)\ma{A}_{J\mu}-(J\leftrightarrow K)]\}.
	\end{split}	
\end{equation}
The notation $\ma{D}_{I}$ will be explained later. The bulk component of the solution can also be found recursively. One can observe that the bulk component is intertwined with the boundary term, so these two equations must be solved in tandem:
\begin{equation}\label{eq:bgcurrentbulkads}
	\begin{split}
		k_I^2\alpha_I=&\frac{\ma{R}}{z}\sum_{I=JK}[\alpha_K(k_I\cdot \ma{A}_J)+\alpha_K(k_K\cdot \ma{A}_J)-\alpha_J(k_I\cdot \ma{A}_K)-\alpha_J(k_J\cdot \ma{A}_K)\\
  &+i(\ma{A}_J\cdot\partial_z\ma{A}_K)-i(\ma{A}_K\cdot\partial_z\ma{A}_J)]\\
		&+\frac{\ma{R}}{z}\mathcal{J}_{Iz}+\frac{\ma{R}^2}{z^2}\sum_{I=JKL}[\alpha_K(\ma{A}_J\cdot\ma{A}_L)-\alpha_L(\ma{A}_J\cdot\ma{A}_K)+(J\leftrightarrow L)],
	\end{split}
\end{equation}
It is important to note that we are not imposing the Lorenz gauge condition, $k_I\cdot \ma{A}_I=0$. The starting point of our recursion is the single point currents which are imposed as $\mathcal{A}_{i\mu}=\phi_{i}(z)\epsilon_{i\mu}$ and $\alpha_{i}=0$ separately for boundary and bulk components. Let $\ma{D}_I^2=\ma{D}_{k_I}^2$ be the d'Alembert operator in AdS:
\begin{equation}
    \ma{D}_{k_I}^2=z^2\partial^2_z+(1-d)z\partial_z-z^2k_I^2,
\end{equation}
and the signle point current $\phi_{i}(z)$ satisfies the Klein-Gordon equation $(\ma{D}_{i}^2-M^{2})\phi_{i}(z)=0$. The interaction current \cite{Armstrong:2022jsa,Tao:2022nqc} between gluons and scalars that are minimally coupled in (A)dS can be expressed as:
\begin{equation}
	\begin{split}
		\mathcal{J}_{I\mu}&=\sum_{I=JK}(-k_{J\mu}\phi_{J}\phi_{K}+k_{K\mu}\phi_{K}\phi_{J})\\
		&~~~~~~+\frac{\ma{R}}{z}\sum_{I=JKL}        \left(\tilde{\mathcal{A}}_{J\mu}\phi_{K}\phi_{L}-2\tilde{\mathcal{A}}_{K\mu}\phi_{J}\phi_{L}+\phi_{J}\tilde{\mathcal{A}}_{L\mu}\phi_{K}\right)\\
		\mathcal{J}_{Iz}&=\sum_{I=JK}(-           
            i\phi_{J}\overleftrightarrow{\partial_{z}}\phi_{K})+\frac{\ma{R}}{z}\sum_{I=JKL}\left(\tilde{\alpha}_{J}\phi_{K}\phi_{L}-2\tilde{\alpha}_{K}\phi_{J}\phi_{L}+\phi_{J}\tilde{\alpha}_{L}\phi_{K}\right).
	\end{split}
\end{equation}
The interaction currents between gluons and minimally coupled scalars in (A)dS have mixed boundary and bulk components. And for cosmological purposes, we specifically consider the case of a minimally coupled scalar that is conformally coupled to gluons. We refer to these scalar currents as $\phi$.

\subsection{The sewing procedure in (A)dS}
Using the BG currents for scalars and gluons discussed above, our goal is to extend the sewing procedure to the (A)dS case, and subsequently build the 1-loop integrands of the Yang-Mills (YM) theory and the scalar theory in (A)dS. Finally, we will explore the unified relation between these theories. The validity of the sewing procedure in (A)dS will not bother us, because the origin of the sewing procedure is simply Feynman rules. In the process of constructing this generalized sewing procedure, one can find the differences between the (A)dS case and the flat case easily.


At the beginning of this subsection, our focus is on constructing the loop integrand for the boundary components of gluons. Similar to the sewing procedure in flat spacetime, we first select a leg from the gluon BG currents and place it in the bulk. An important note is that, for the legs we select and place in the bulk, the integral of the radial component for vertices connected by them must be removed before the sewing procedure and restored afterward. Therefore, the currents we consider must have two types for general terms and one type for tadpole terms. For the boundary component of the gluon BG currents, we have the following (in the following, we assume the (A)dS radius $\ma{R}=1$):

\begin{equation}\label{eq:BGcurrentboundaryleg}
	\begin{split}
		\frac{1}{z^2}(\mathcal{D}_{I'}^2+d-1)\ma{A}_{I'\mu}&=ik_{lI\mu}(\partial_z+\frac{2-d}{z})\alpha_{lI}+\frac{1}{z}\sum_{I'=lI}\{(k_{I\mu}\alpha_I+2i\partial_z\ma{A}_{I\mu})\alpha_l\\
		&+k_{I\mu}(\ma{A}_l\cdot\ma{A}_I)+\ma{A}_{I\mu}[i(\partial_z-\frac{d}{z})\alpha_l-k_{lI}\cdot\ma{A}_l-k_{I}\cdot\ma{A}_{l}]-(l\leftrightarrow I)\}\\
		&+\frac{1}{z^2}\sum_{I'=lJK}\{[\alpha_l\alpha_J\ma{A}_{K\mu}+(\ma{A}_l\cdot\ma{A}_J)\ma{A}_{K\mu}-(J\leftrightarrow K)]\\
		&+[\alpha_J\alpha_K\ma{A}_{l\mu}+(\ma{A}_J\cdot\ma{A}_K)\ma{A}_{l\mu}-(l\leftrightarrow J)]\}\\
		&+\frac{1}{z}\sum_{I'=lJK}\{(k_{K\mu}\alpha_K+2i\partial_z\ma{A}_{K\mu})\alpha_{lJ}+k_{K\mu}(\ma{A}_{lJ}\cdot\ma{A}_K)\\
		&+\ma{A}_{K\mu}[i(\partial_z-\frac{d}{z})\alpha_{lJ}-k_{lJK}\cdot\ma{A}_{lJ}-k_{K}\cdot\ma{A}_{lJ}]-(lJ\leftrightarrow K)\}\\
		&+\frac{1}{z^2}\sum_{I'=lJKL}\{[\alpha_{lJ}\alpha_K\ma{A}_{L\mu}+(\ma{A}_{lJ}\cdot\ma{A}_K)\ma{A}_{L\mu}-(K\leftrightarrow L)]\\
		&+[\alpha_K\alpha_L\ma{A}_{lJ\mu}+(\ma{A}_K\cdot\ma{A}_L)\ma{A}_{lJ\mu}-(lJ\leftrightarrow K)]\}-\frac{1}{z}\mathcal{J}_{lI\mu}.
	\end{split}
\end{equation}
The terms in the first four lines, similar to the pre-integrand in Minkowski spacetime, contribute to the tadpole diagrams. However, unlike in flat spacetime where the tadpole diagrams vanish after dimensional regularization, in (A)dS spacetime, the propagator has an effective mass, meaning that the tadpoles may also contribute.

Similar to the flat case, we can observe that only the terms resulting from two boundary gluon legs sewn together will possibly survive from $\ma{T}[ln]$ prior to the sewing procedure. Therefore, we will only consider this scenario in our subsequent discussion. We represent the legs selected from boundary gluon currents as $\ma{A}_{lI\nu}=\ma{A}_{l}^{\mu}\ma{J}_{\mu\nu}$, and those originating from bulk gluon currents as $\alpha_{lI}=\ma{A}_{l}^{\mu}\ma{K}_{I\mu}$.


For the case that the two legs to be sewed are both the boundary component $\ma{A}_{l\mu}$ (which means that $\alpha_{l}=0$), the boundary component current can be written as:
\begin{equation}\label{glurec}
	\begin{split}
		\frac{1}{z^2}(\mathcal{D}_{I'}^2+d-1)\ma{A}_l^{\nu}\ma{J}_{I\mu\nu}&=ik_{lI\mu}(\partial_z+\frac{2-d}{z})\ma{A}_{l}^{\nu}\ma{K}_{I\nu}-\frac{1}{z}\mathcal{J}_{lI\mu}+\frac{1}{z}\{-2i\partial_z\ma{A}_{l\mu}\alpha_I\\
		&~~~~~~+(k_{I\mu}-k_{l\mu})(\ma{A}_l\cdot\ma{A}_I)-\ma{A}_{I\mu}(k_{lI}\cdot\ma{A}_l)-\ma{A}_{I\mu}(k_{I}\cdot\ma{A}_l)\\
  		&~~~~~~-\ma{A}_{l\mu}[i(\partial_z-\frac{d}{z})\alpha_I-k_{lI}\cdot\ma{A}_I-k_{l}\cdot\ma{A}_I]\}\\
		&+\frac{1}{z^2}\sum_{I'=lJK}\{[(\ma{A}_l\cdot\ma{A}_J)\ma{A}_{K\mu}-(J\leftrightarrow K)]\\
		&~~~~~~+[\alpha_J\alpha_K\ma{A}_{l\mu}+(\ma{A}_J\cdot\ma{A}_K)\ma{A}_{l\mu}-(\ma{A}_l\cdot\ma{A}_K)\ma{A}_{J\mu}]\}\\
		&+\frac{1}{z}\sum_{I'=lJK}\{(k_{K\mu}\alpha_K+2i\partial_z\ma{A}_{K\mu})\ma{A}_l^{\nu}\ma{K}_{J\nu}\\
		&~~~~~~-(k_{lJ\mu}\ma{A}_l^{\nu}\ma{K}_{J\nu}-2i\partial_z(\ma{A}_l^{\nu}\ma{J}_{J\mu\nu}))\alpha_K+\ma{A}_l^{\nu}k_{K\mu}(\ma{J}_{J\nu}\cdot\ma{A}_K)\\
		&~~~~~~+\ma{A}_{K\mu}[i(\partial_z-\frac{d}{z})\ma{A}_{l}^{\nu}\ma{K}_{J\nu}-k_{lJK}\cdot\ma{A}_l^{\nu}\ma{J}_{J\nu}-k_{K}\cdot\ma{A}_l^{\nu}\ma{J}_{J\nu}]\\
		&~~~~~~-k_{lJ\mu}(\ma{A}_K\cdot\ma{J}_{J\nu}\ma{A}_l^{\nu})\\
		&~~~~~~-\ma{A}_l^{\nu}\ma{J}_{J\mu\nu}[i(\partial_z-\frac{d}{z})\alpha_K-k_{lJK}\cdot\ma{A}_K-k_{lJ}\cdot\ma{A}_K]\}\\
		&+\frac{1}{z^2}\sum_{I'=lJKL}\{[\ma{A}_{l}^{\nu}\ma{K}_{J\nu}\alpha_K\ma{A}_{L\mu}+\ma{A}_l^{\nu}(\ma{J}_{J\nu}\cdot\ma{A}_K)\ma{A}_{L\mu}\\
		&~~~~~~~~~~~~~~~-\ma{A}_{l}^{\nu}\ma{K}_{J\nu}\alpha_L\ma{A}_{K\mu}-\ma{A}_l^{\nu}(\ma{J}_{J\nu}\cdot\ma{A}_L)\ma{A}_{K\mu}]\\
		&~~~~~~+[\alpha_K\alpha_L\ma{A}_l^{\nu}\ma{J}_{J\mu\nu}+(\ma{A}_K\cdot\ma{A}_L)\ma{A}_l^{\nu}\ma{J}_{J\mu\nu}\\
		&~~~~~~-\ma{A}_l^{\nu}\ma{K}_{J\nu}\alpha_L\ma{A}_{K\mu}-\ma{A}_l^{\nu}(\ma{J}_{J\nu}\cdot\ma{A}_L)\ma{A}_{K\mu}]\}.
	\end{split}
\end{equation}
%
It is worth noting that the presence of terms involving $\partial_{z}\ma{A}_{l}^{\mu}$ may seem perplexing. In particular, when gluing together $\partial_{z}\ma{A}_{l}^{\mu}$ and $\ma{A}_{n}^{\mu}$ in the sewing procedure to construct the loop propagator, the meaning of this operation needs further clarification. However, the correlation between these partial derivatives can be understood through integration by parts, as the $z$ integral is carried out for each bulk vertex.

The next step involves sewing the two legs we selected together and replacing the corresponding single-point currents with the two-point correlation function. Then we restore the cyclic completion by hand, remove the redundant terms, and finally obtain the loop integrand in (A)dS. To be precise, after the following replacement:
\ie
\ma{A}_{l}^{\mu}(z')\ma{A}_{n}^{\nu}(z)&\to\eta^{\mu\nu}\la\phi_{l}(z')\phi_{n}(z)\ra\\
(\partial\ma{A}_{l}^{\mu})(z')\ma{A}_{n}^{\nu}(z)&\to\eta^{\mu\nu}\la(\partial\phi_{l})(z')\phi_{n}(z)\ra
\fe
%
By modifying the deconcatenation sum in the same way as in the flat case, we can obtain $\eta^{\mu\nu}\tilde{\ma{J}}_{\mu\nu}(z,z')$ from $\ma{A}^{\mu}_{n}(\mathcal{D}_{I'}^2+d-1)\ma{A}_l^{\nu}\ma{J}_{\mu\nu}(z,z')$. The pre-integrand can then be defined as $I_{\text{1-loop}}=\eta_{\mu\nu}\tilde{\ma{J}}^{\mu\nu}$. In this context, the correlation function $\la\phi_{n}(z)\phi_{l}(z')\ra$ is the same as the correlation function $G_{k_{l}}(z,z')$, which denotes the momentum of the internal line resulting from the sewing of the legs $l$ and $n$.


The aforementioned steps are expected to be straightforward. For the subsequent unifying relation, we simply remove the terms that cannot contain terms proportional to $\eta_{\mu\nu}$ (where we have employed $k_{I}=0$).
\ie
\tilde{\ma{J}}^{\text{stripped}}_{I\mu\nu}&=-z\{2i\alpha_{I}\eta_{\mu\nu}\la\phi_{n}\partial_{z}\phi_{l}\ra+\eta_{\mu\nu}\la\phi_{n}\phi_{l}\ra[i(\partial_z-\frac{d}{z})\alpha_I-2(k_{l}\cdot\ma{A}_I)]\}\\
&+\sum_{I=JK}\la\phi_{n}\phi_{l}\ra(\ma{A}_J\cdot\ma{A}_K+\alpha_{J}\alpha_{K})\eta_{\mu\nu}\\
&+z\sum_{I=JK}\{-2i\la\phi_{n}\phi_{l}\ra(\partial_z\ma{J}_{J\mu\nu})\alpha_K+\la\phi_{n}\phi_{l}\ra\ma{J}_{J\mu\nu}[i(\partial_z-\frac{d}{z})\alpha_K-k_{lJK}\cdot\ma{A}_K-k_{lJ}\cdot\ma{A}_K]\}\\
&+\la\phi_{n}\phi_{l}\ra\sum_{I=JKL}[\alpha_K\alpha_L\ma{J}_{J\mu\nu}+(\ma{A}_K\cdot\ma{A}_L)\ma{J}_{J\mu\nu}].
\fe


To construct the loop integrand for a conformally coupled scalar in (A)dS, we can follow the same steps as for the gauge field. Let us start by writing down the tree-level BG currents for the scalar field in (A)dS:
\begin{equation}\label{eq:scalarrecur}
    \begin{split}
        \frac{1}{z^2}(\ma{D}_I^2+d-1)\phi_I&=\frac{1}{z}\sum_{I=JK}[\phi_J(k_I\cdot\tilde{\ma{A}}_K)+\phi_J(k_J\cdot\tilde{\ma{A}}_K)\\
        &~~~~~~~~-i(\phi_J\partial_z\tilde{\alpha}_K+2\tilde{\alpha}_K\partial_z\phi_J-\frac{d}{z}\phi_J\tilde{\alpha}_K)-(J\leftrightarrow K)]\\
        &~~~+\frac{1}{z^2}\sum_{I=JKL}[(\tilde{\ma{A}}_J\cdot\tilde{\ma{A}}_K)\phi_L-(\tilde{\ma{A}}_J\cdot\tilde{\ma{A}}_L)\phi_K\\
        &~~~~~~~~~~~~~~~~~~+\tilde{\alpha}_J\tilde{\alpha}_K\phi_L-\tilde{\alpha}_J\tilde{\alpha}_L\phi_K+(J\leftrightarrow L)].
    \end{split}
\end{equation}

The BG current for the scalar field is denoted by $\phi$, which like the gauge fields BG currents in curved spacetime, can be decomposed into bulk and boundary components. In the scalar theory, the bulk component of the BG currents is given by $\tilde{\alpha}_I$, while the boundary component is denoted by $\tilde{\ma{A}}_I$.


Next, we select one leg and consider it as being on the bulk, which we denote as $\phi_l$. It should be noted that this leg should not be confused with the one-point scalar BG current. In what follows, we assume that all the legs in $I$ correspond to boundary components of the gluons, so we have $\tilde{\ma{A}}_{J\mu}=\ma{A}_{J\mu}$ if $J\subset I$.
\ie
\frac{1}{z^2}(\ma{D}_I^2+d-1)\phi_{l}\Phi_{I}&=\frac{1}{z}[\phi_{l}(k_{lI}\cdot\tilde{\ma{A}}_I)+\phi_{l}(k_{l}\cdot\tilde{\ma{A}}_I)-i(\phi_l\partial_z\tilde{\alpha}_I+2\tilde{\alpha}_I\partial_z\phi_l-\frac{d}{z}\phi_l\tilde{\alpha}_I)]\\
&+\frac{1}{z^2}\sum_{I=KL}[(\tilde{\ma{A}}_L\cdot\tilde{\ma{A}}_K)\phi_l+\tilde{\alpha}_L\tilde{\alpha}_K\phi_l]\\
&+\frac{1}{z}\sum_{I=JK}[\phi_{l}\Phi_{J}(k_{lI}\cdot\tilde{\ma{A}}_K)+\phi_{l}\Phi_{J}(k_{lJ}\cdot\tilde{\ma{A}}_K)\\
&-i(\Phi_{J}\phi_{l}\partial_z\tilde{\alpha}_K+2\tilde{\alpha}_K\partial_z(\phi_{l}\Phi_{J})-\frac{d}{z}\phi_{l}\Phi_{J}\tilde{\alpha}_K)]\\
&+\frac{1}{z^2}\sum_{I=JKL}[(\tilde{\ma{A}}_{L}\cdot\tilde{\ma{A}}_K)\Phi_J\phi_{l}+\tilde{\alpha}_{L}\tilde{\alpha}_K\Phi_J\phi_{l}].
\fe
%
Once again, the terms in the first two lines will contribute to the scalar tadpole diagrams and will not be considered in this work. The next step involves gluing the two bulk legs together and replacing the two formal single-point BG currents with a propagator. This allows us to obtain the loop integrand with a scalar loop:
\ie
\tilde{\Phi}^s_{I}&=z[2\la\phi_{n}\phi_{l}\ra(k_l\cdot\tilde{\ma{A}}_I)-i(\la\phi_{n}\phi_{l}\ra\partial_z\tilde{\alpha}_I+2\tilde{\alpha}_I\la\phi_{n}\partial_z\phi_l\ra-\la\phi_{n}\phi_{l}\ra\frac{d}{z}\tilde{\alpha}_I)]\\
&+\sum_{I=KL}[\la\phi_{n}\phi_{l}\ra(\tilde{\ma{A}}_L\cdot\tilde{\ma{A}}_K)+\la\phi_{n}\phi_{l}\ra\tilde{\alpha}_L\tilde{\alpha}_K]\\
&+z\sum_{I=JK}[2\la\phi_{n}\phi_{l}\ra\tilde{\Phi}^{s}_{J}(k_l\cdot\tilde{\ma{A}}_K)-i(\la\phi_{n}\phi_{l}\ra\tilde{\Phi}^s_{J}\partial_z\tilde{\alpha}_K\\
&+2\la\phi_{n}\phi_{l}\ra\tilde{\alpha}_K\partial_z\tilde{\Phi}^{s}_{J}-\la\phi_{n}\phi_{l}\ra\frac{d}{z}\tilde{\Phi}^s_{J}\tilde{\alpha}_K)]\\
&+\sum_{I=JKL}[\la\phi_{n}\phi_{l}\ra(\tilde{\ma{A}}_{L}\cdot\tilde{\ma{A}}_K)\tilde{\Phi}^s_J+\la\phi_{n}\phi_{l}\ra\tilde{\alpha}_{L}\tilde{\alpha}_K\tilde{\Phi}^s_J].
\fe


Therefore, we can express the unifying relation in (A)dS for the one-loop integrand as follows:
\ie\label{lo}
		\hat{\ma{D}}\tilde{\ma{J}}_{I\mu\nu}=\tilde{\Phi}^{s}_{I}
\fe
Here, the operator $\hat{\ma{D}}$ selects the terms proportional to $\eta_{\mu\nu}$ in $\tilde{\ma{J}}_{\mu\nu}$. The proof of these relations is the same as in the flat case, which is based on the proof of the tree-level one in Section \ref{sec:flatproof}. The reason for using the operator $\hat{\ma{D}}$ rather than $\ma{D}=\frac{\partial}{\partial d}$ as in the flat case is that there will be also some $d$ coming from the (A)dS measure $\sqrt{-g}=z^{-d-1}$. However, only the $d$ that comes from the sewing procedure, i.e. comes from $\eta^{\mu\nu}\eta_{\mu\nu}$, contributes to the unifying relation we want to prove. Hence we must choose the operator $\hat{\ma{D}}$ in order to avoid the interference coming from the (A)dS metric.



After integrating over the radial coordinate, we obtain the following expression for the loop integrand:
\begin{align}
	I_{\text{1-loop}}(P) = \int \frac{dz}{z^{d+1}}\tilde{\Phi}_{P}^{\text{tadpole}}(z) + \int \frac{dz}{z^{d+1}}\int \frac{dz'}{z'^{d+1}}\tilde{\Phi}^{\text{non-tadpole}}_{P}(z,z').
\end{align}
This equation establishes a relationship between the loop integrands and the currents in the (A)dS case.

\subsection{2-pt one loop correlation function}
In this section, we will provide an example to validate the relation discussed earlier. Specifically, we will compute the two-point bubble diagram in the $\xi=1$ gauge and refer to \cite{Albayrak:2018tam} for details.
\begin{figure}[H]
	\centering
	\begin{tikzpicture}[line width=1pt,scale=1.5]
		\draw[photon] (2.5,0)--(3.6,0);
		\draw[photon,fill=white] (4,0) circle (.5cm);
		\draw[photon] (4.5,0)--(5.6,0);
		\node at (2.1,0) {$\mu;a$};
        \node at (2.1,0.3) {$\stackrel{k_1}{\rightarrow}$};
		\node at (3.2,0.3) {$\sigma;d$};
		\node at (3.2,-0.3) {$\rho;c$};
		\node at (4,0.8) {$\stackrel{q}{\rightarrow}$};
		\node at (4.7,0.3) {$\lambda;f$};
		\node at (4.7,-0.3) {$\tau;e$};
		\node at (4,-0.8) {$\stackrel{p}{\leftarrow}$};
		\node at (6,0) {$\nu;b$};
        \node at (6,0.3) {$\stackrel{k_2}{\leftarrow}$};
	\end{tikzpicture}
	\caption{2-pt one loop gluon correlation function}
\end{figure}
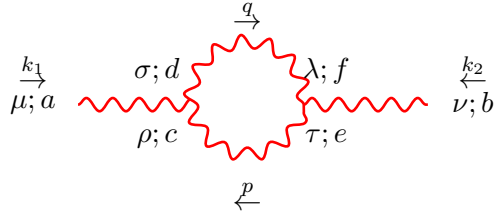

\subsubsection{Feynman rule perspective}
First, let us calculate the 1-loop gluon propagator with a boundary component of a gluon running in the loop (since the radial component loop has no contribution to the unifying relation) using the standard Feynman rule. It is not difficult to derive the Feynman rule for gluons in (A)dS. For the 3-point vertex, we have:
\begin{equation}
	\begin{tikzpicture}[line width=1pt,scale=1.5]
		\begin{scope}[shift={(0,0)}]
			\draw[photon] (90:0) -- (90:1cm);
			\draw[photon] (-30:0) -- (-30:1cm);
			\draw[photon] (210:0) -- (210:1cm);
			\node at (90:1.2cm) {$\mu,a$};
			\node at (210:1.2cm) {$\nu,b$};
			\node at (-30:1.2cm) {$\rho,c$};
			\draw[fermion] (0.2,0.8) -- (0.2,0.2);
			\begin{scope}[shift={(0,-0.2)}]
			    \draw[fermionbar] (-30:0.2) -- (-30:0.8cm);
			    \node at (0.3,-0.45) {$q$};
			\end{scope}
			\begin{scope}[shift={(0,0.2)}]
			    \draw[fermionbar] (210:0.2) -- (210:0.8cm);
			    \node at (-0.5,0.05) {$p$};
			\end{scope}
			\node at (0.4cm,0.5cm) {$k$};
		\end{scope}
		\begin{scope}[shift={(3,0)}]
			\node at (1.5,0.0) {$\displaystyle=g\int_0^{\infty}z^{-d}f^{abc}[\eta^{\mu\nu}(k-p)^{\rho}+\eta^{\nu\rho}(p-q)^{\mu}+\eta^{\rho\mu}(q-k)^{\nu}]$.};
		\end{scope}	
	\end{tikzpicture}
\end{equation}

Utilizing the Feynman rule mentioned earlier, we can express the one-loop 2-point correlation function as follows:
\begin{equation}
	\begin{split}
		\langle \epsilon_{1}\cdot A^{a}_{1}\epsilon_{2}\cdot A_{2}^{b}\rangle=&g^2\int dz_1dz_2 z_1^{-d}z_2^{-d}\epsilon_{1\mu}\epsilon_{2\nu}\phi(z_{1})\phi(z_{2})f^{acd}f^{bfe}\\
		&\times\int d^dq\big[\eta^{\mu\rho}(k_1-p)^{\sigma}+\eta^{\rho\sigma}(p+q)^{\mu}-\eta^{\sigma\mu}(q+k_1)^{\rho}\big]\\
		&\times\big[-\eta^{\nu\tau}(k_2+p)^{\lambda}+\eta^{\lambda\tau}(q+p)^{\nu}+\eta^{\lambda\nu}(-q+k_2)^{\tau}\big]G_{\sigma\lambda}^{df}(z_1,z_2;q)G_{\rho\tau}^{ce}(z_1,z_2;p),
	\end{split}
\end{equation}
where $G_{\sigma\lambda}^{df}(z_1,z_2;q)=G_{q}(z_{1},z_{2})\delta^{df}\eta_{\sigma\lambda}$, and $G_{q}(z_{1},z_{2})=\la\phi(z_{1})\phi(z_{2})\ra$ with momentum $q$, and the loop momentum $p=q-k_{1}$ also contributes to the integral with respect to $q$. It's worth noting that the color factor $f^{acd}f^{bdc}$ can be expressed as $-\tr(T^{i}[T^{a},[T^{b},T^{i}]])$. Thus, the partial correlation function becomes
\begin{equation}
	\begin{split}
		\langle \epsilon_{1}\cdot A_{1}\epsilon_{2}\cdot A_{2}\rangle_{\text{partial}}=&-g^2\int dz_1dz_2 z_1^{-d}z_2^{-d}\int d^dq\big[-6(p\cdot \epsilon_{1})(p\cdot \epsilon_{2})+4d(p\cdot \epsilon_{1})(p\cdot \epsilon_{2})\\
		&+(-5k_{1}\cdot k_{2}+2k_{1}\cdot p+2p^{2})(\epsilon_{1}\cdot \epsilon_{2})\big]G_{q}(z_1,z_2)G_{p}(z_1,z_2)\phi(z_{1})\phi(z_{2}).
	\end{split}
\end{equation}

Next, let us consider the scalar theory. We can also compute the 1-loop 2-point gluon correlation with a scalar running in the loop. To do so, we need to write down the interaction terms between gluons and scalars in the Lagrangian. It is worth noting that we are working in the $\xi=1$ gauge in this section.

\ie\label{eq:interactionlag}
\tr(g^{mn}\mb{A}_{m}\mb{J}_{n})&=\tr(z^{2}\mb{A}_{z}[i\partial_{z}\phi+[\mb{A}_{z},\phi],\phi]+z^{2}\eta^{\mu\nu}\mb{A}_{\mu}[i\partial_{\nu}\phi+[\mb{A}_{\nu},\phi],\phi])\\
&=-zf^{abc}\alpha^{c}\partial_{z}\phi^{a}\phi^{b}-f^{abc}f^{cde}\alpha^{a}\phi^{b}\phi^{d}\alpha^{e}-zf^{abc}\ma{A}^{c\mu}\partial_{\mu}\phi^{a}\phi^{b}-f^{abc}f^{cde}\ma{A}_{\mu}^{a}\ma{A}^{e\mu}\phi^{b}\phi^{d}.
\fe
Then, we can write down the Feynman rule for vertex among $\phi$ and $A$ according to the Lagrangian \eqref{eq:interactionlag} above:
\begin{equation}
	\begin{tikzpicture}[line width=1pt,scale=1.5]
		\begin{scope}[shift={(0,0)}]
			\draw[scalarnoarrow] (90:0) -- (90:1cm);
			\draw[scalarnoarrow] (-30:0) -- (-30:1cm);
			\draw[photon] (210:0) -- (210:1cm);
			\node at (90:1.2cm) {$a$};
			\node at (210:1.2cm) {$\mu,c$};
			\node at (-30:1.2cm) {$b$};
			\draw[fermion] (0.2,0.8) -- (0.2,0.2);
			\begin{scope}[shift={(0,-0.2)}]
			    \draw[fermionbar] (-30:0.2) -- (-30:0.8cm);
			    \node at (0.3,-0.45) {$q$};
			\end{scope}
			\begin{scope}[shift={(0,0.2)}]
			    \draw[fermionbar] (210:0.2) -- (210:0.8cm);
			    \node at (-0.5,0.05) {$p$};
			\end{scope}
			\node at (0.4cm,0.5cm) {$k$};
		\end{scope}
		\begin{scope}[shift={(3,0)}]
			\node at (1.5,0.0) {$\displaystyle=g\int_0^{\infty}z^{-d}f^{abc}(q_{\mu}-k_{\mu})$.};
		\end{scope}	
	\end{tikzpicture}
\end{equation}
Therefore, for the correlation function $\langle AA\rangle$ with a scalar loop, the integrand can be expressed as follows:
\ie\label{eq:2ptgluonscalarloop}
\ma{I}_{AA,\text{scalar}}=-g^{2}\int dz_1 dz_2 z_1^{-d}z_2^{-d} f^{acd}f^{bcd}(p+q)\cdot\epsilon_{1}\phi_{1}(z_1)G_{p}(z_1,z_2)G_{q}(z_1,z_2)(p+q)\cdot\epsilon_{2}\phi_{2}(z_2),
\fe
whereas the partial correlation function is given by:
\ie
\ma{I}^{\text{partial}}_{AA,\text{scalar}}=-g^{2}\int dz_1 dz_2 z_1^{-d}z_2^{-d} (p+q)\cdot\epsilon_{1}\phi_{1}(z_1)G_{p}(z_1,z_2)G_{q}(z_1,z_2)(p+q)\cdot\epsilon_{2}\phi_{2}(z_2)
\fe




\subsubsection{Sewing procedure perspective}
We can also demonstrate that the correct loop integrands can be obtained by using BG currents (we have set the YM coupling constant $g$ to be 1). In the diagram we are considering, a bubble diagram, the tadpoles do not contribute. The only terms that contribute are:
\ie
	&z\sum_{I'=lJ|K}\{(k_{K\mu}\alpha_K+2i\partial_z\ma{A}_{K\mu})\ma{A}_l^{\nu}\ma{K}_{J\nu}-(k_{lJ\mu}\ma{A}_l^{\nu}\ma{K}_{J\nu}-2i\partial_z(\ma{A}_l^{\nu}\ma{J}_{J\mu\nu}))\alpha_K\\
		&+\ma{A}_l^{\nu}k_{K\mu}(\ma{J}_{J\nu}\cdot\ma{A}_K)+\ma{A}_{K\mu}[i(\partial_z-\frac{d}{z})\ma{A}_{l}^{\nu}\ma{K}_{J\nu}-k_{lJK}\cdot\ma{A}_l^{\nu}\ma{J}_{J\nu}-k_{K}\cdot\ma{A}_l^{\nu}\ma{J}_{J\nu}]\\
		&-k_{lJ\mu}(\ma{A}_K\cdot\ma{J}_{J\nu}\ma{A}_l^{\nu})-\ma{A}_l^{\nu}\ma{J}_{J\mu\nu}[i(\partial_z-\frac{d}{z})\alpha_K-k_{lJK}\cdot\ma{A}_K-k_{lJ}\cdot\ma{A}_K]\}.
\fe
In the case of the 2-point function, we have $I=12$, which implies that legs 1 and 2 are boundary legs. Therefore, we have:
\ie\label{aaa}
	&z\{\ma{A}_l^{\nu}k_{2\mu}(\ma{J}_{1\nu}\cdot\ma{A}_2)+\ma{A}_{2\mu}[-k_{l12}\cdot\ma{A}_l^{\nu}\ma{J}_{1\nu}-k_{2}\cdot\ma{A}_l^{\nu}\ma{J}_{1\nu}]\\
		&-k_{l1\mu}(\ma{A}_2\cdot\ma{J}_{1\nu}\ma{A}_l^{\nu})+\ma{A}_l^{\nu}\ma{J}_{1\mu\nu}[k_{l12}\cdot\ma{A}_2+k_{l1}\cdot\ma{A}_2]\}.
\fe
The 2-point currents in the YM theory can be obtained easily from the recursions. For pure boundary gluon diagrams, all $\alpha$ currents are zero, so we have:
\ie\label{eq:2ptgluoncurrent}
(\mathcal{D}_{l1}^2+d-1)\mathcal{A}_{l1\mu}&=z[(k_{1\mu}-k_{l\mu})(\mathcal{A}_{l}\cdot\mathcal{A}_{1})+2\ma{A}_{l\mu}(k_{l}\cdot\ma{A}_{1})-\mathcal{A}_{1\mu}(k_{1}\cdot\mathcal{A}_{l})-\ma{A}_{1\mu}(k_{l1}\cdot\ma{A}_{l})]\\
&=z\ma{A}_{l}^{\nu}[(k_{1\mu}-k_{l\mu})\mathcal{A}_{1\nu}+2\eta_{\mu\nu}(k_{l}\cdot\ma{A}_{1})-\mathcal{A}_{1\mu}k_{1\nu}-\ma{A}_{1\mu}k_{l1\nu}].
\fe
%
It is important to note that there is no radial coordinate integration in the expression \eqref{aaa}. Therefore, we can directly set $1/(\mathcal{D}_{l1}^2+d-1)$ to be equal to the bulk-to-bulk propagator $G_{q}(z,z')$, without integrating over $z'$. In this case, we have $k_{l\mu}=p_{\mu}$ and $q_{\mu}=k_{l1\mu}$, since there are two legs in the bulk and two vertices in the diagram. Next, we contract \eqref{aaa} with $\ma{A}_{n\mu}$ and apply the sewing procedure.
\ie
\ma{A}_{l}^{\mu}(z')\ma{A}_{n}^{\nu}(z)&\to\eta^{\mu\nu}\la\phi_{l}(z')\phi_{n}(z)\ra=\eta_{\mu\nu}G_{p}(z,z')\\
(\partial\ma{A}_{l}^{\mu})(z')\ma{A}_{n}^{\nu}(z)&\to\eta^{\mu\nu}\la(\partial\phi_{l})(z')\phi_{n}(z)\ra=\eta_{\mu\nu}\partial_{z'}G_{p}(z,z').
\fe
After performing the sewing procedure, we obtain the one-loop integrand for the correlator without the radial integral $\tilde{I}_{\text{1-loop}}$, which can be expressed as follows:
\ie
	\tilde{I}_{\text{1-loop}}=&zz'G_{p}(z,z')G_{q}(z,z')\{[(k_{1}-k_{l})\cdot\ma{A}_{2}\mathcal{A}_{1}\cdot k_{2}+2k_{2}\cdot\ma{A}_{2}(k_{l}\cdot\ma{A}_{1})-\mathcal{A}_{1}\cdot\ma{A}_{2}(k_{1}+k_{l1})\cdot k_{2}]\\
 &-[(k_{1}-k_{l})\cdot (k_{2}+k_{l})\mathcal{A}_{1}\cdot\ma{A}_{2}+2(k_{2}+k_{l})\cdot\ma{A}_{2}(k_{l}\cdot\ma{A}_{1})-\mathcal{A}_{1}\cdot (k_{2}+k_{l})(k_{1}+k_{l1})\cdot\ma{A}_{2}]\\
		&-[(k_{1}-k_{l})\cdot\ma{A}_{2}\mathcal{A}_{1}\cdot k_{l1}+2k_{l1}\cdot\ma{A}_{2}(k_{l}\cdot\ma{A}_{1})-\mathcal{A}_{1}\cdot\ma{A}_{2}(k_{1}+k_{l1})\cdot k_{l1}]\\
  &+2[(k_{1}-k_{l})\cdot\mathcal{A}_{1}+2d(k_{l}\cdot\ma{A}_{1})-\mathcal{A}_{1}\cdot (k_{1}+k_{l1})](k_{l}\cdot\ma{A}_2)\}\\
  =&zz'G_{p}(z,z')G_{q}(z,z')[-6(k_{l}\cdot \ma{A}_{1})(k_{l}\cdot \ma{A}_{2})+4d(k_{l}\cdot \ma{A}_{1})(k_{l}\cdot \ma{A}_{2})\\
  &~~~+(-5k_{1}\cdot k_{2}+2k_{1}\cdot k_{l}+2k_{l}^{2})(\ma{A}_{1}\cdot \ma{A}_{2})].
\fe

It is worth noting that the previous expression for the one-loop integrand, which lacked the radial integral, did not explicitly write the spacetime dependence of $\ma{A}_{1\mu}$ and $\ma{A}_{2\mu}$. In order to restore the radial integral, we need to consider this dependence. In particular, $\ma{A}_{1\mu}$ depends only on $z$, while $\ma{A}_{2\mu}$ depends only on $z'$. Therefore, we can reintroduce the radial integral and write the one-loop integrand as:
%
\ie
I_{\text{1-loop}}=&\int\frac{dz}{z^{d}}\int\frac{dz'}{z'^{d}}G_{p}(z,z')G_{q}(z,z')[-6(k_{l}\cdot \ma{A}_{1})(k_{l}\cdot \ma{A}_{2})+4d(k_{l}\cdot \ma{A}_{1})(k_{l}\cdot \ma{A}_{2})\\
&+(-5k_{1}\cdot k_{2}+2k_{1}\cdot k_{l}+2k_{l}^{2})(\ma{A}_{1}\cdot \ma{A}_{2})],
\fe
The expression for the 1-loop integrand (scalar loop) of the scalar theory, obtained from the unifying relation \eqref{lo}, is given by:
\ie
I_{\text{1-loop}}^{s}=4\int\frac{dz}{z^{d}}\int\frac{dz'}{z'^{d}}G_{p}(z,z')G_{q}(z,z')(k_{l}\cdot \ma{A}_{1})(k_{l}\cdot \ma{A}_{2})
\fe
%

The result obtained from the unifying relation for the 1-loop integrand (scalar loop) of the scalar theory is identical to the result obtained from Feynman rules, except for an overall minus sign. This minus sign comes from the difference between the propagators in Feynman rules and the propagators in BG currents. More precisely, in the flat case, the propagators in Feynman rules look like $i/p^2$, while $1/p^2$ in BG currents. Such an argument is also valid for the (A)dS case since we can take the flat limit to obtain the flat propagators from the bulk-to-bulk (A)dS propagators. Therefore, we have successfully verified the sewing procedure and the unifying relation at the 1-loop level in this case.

\section{Conclusion and Outlook}



In this work, we utilize BG currents to construct differential operators for 1-loop integrands and demonstrate the corresponding unifying relations at the 1-loop level in both flat and (A)dS spacetime. Specifically, we focus on the unifying relations between YMS and YM theory \eqref{eq:unifyingYMSandYM} as well as their corollary, where the gluons are minimally coupled to scalars \eqref{eq:unifyingYMSandYM}. We also investigate the (A)dS case.


At the outset of our work, we provided a brief review of some differential operators and unifying relations. We then outlined the BG currents approach to the unifying relation \eqref{eq:unifyingminicouplescalarandYM} between YM and minimally coupled scalars at the tree level. To extend these results to the 1-loop level, we introduced the sewing procedure in Section \ref{sec:flat} to construct the 1-loop integrand from BG currents. We expect that the unifying relations \eqref{ymsuni} at the 1-loop level can be obtained from the tree-level amplitudes during the sewing procedure. In Section \ref{sec:flatproof}, we presented the explicit form of the differential operators and proved the corresponding unifying relations \eqref{ymsuni} between YM and minimally coupled scalars at the 1-loop level. Additionally, we demonstrated the differential operators and unifying relations between YMS and YM theory \eqref{eq:ymsunifyingloop}. The proof we give here, which is based on BG recursion, means that we find a semi-on-shell way to the unifying relation which is used to be proved by on-shell factorization.


Constructing differential operators in (A)dS can be more complex than in flat spacetime. In flat spacetime, the metric contraction, when sewing two legs to a loop, is the only source of the spacetime dimension $d$, and we can extract the relevant contribution only using the spacetime dimension differential operator. However, in (A)dS, the spacetime dimension $d$ in BG currents can also come from the (A)dS metric, leading to some ambiguities when proving the unifying relations. To avoid these ambiguities, in Section \ref{sec:ads}, we construct the differential operator $\hat{\ma{D}}$ instead of using the naive spacetime dimension differential operator $\ma{D}$. Additionally, we explicitly calculate the gluon 1-loop integrand using Feynman rules and confirm that the results match the expression obtained from the sewing procedure.

The success of the generalization of the unifying relation discussed in this paper shows that correlators in (A)dS spacetime have some features similar to the flat case. This will deepen our understanding of (A)dS correlators. There remain several open questions that we hope to address in future work. One question is whether there exist nontrivial relations involving the graviton BG currents \cite{Gomez:2021shh}. Furthermore, could the unifying relations involving effective theories, such as the Dirac-Born-Infeld theory and the special Galileon theory, be generalized to the (A)dS spacetime? Another interesting direction would be to extend our discussion to the double copy formalism \cite{Lee:2022fgr,Albayrak:2020fyp,Li:2022tby,Cho:2021nim,Farrow:2018yni,Armstrong:2020woi}. We look forward to exploring these topics in the future.

\section{Acknowledgement}
We would like to thank Yi-Jian Du and Kang Zhou for useful comments on the draft. QC is partly supported by National Key R\&D Program of China (Grant No. 2017YFA0402204). YT is partly supported by National Key R\&D Program of China (NO. 2020YFA0713000).

\bibliographystyle{JHEP}
\bibliography{loop}	

\end{document}